\documentclass[prd,aps,12pt,%showpacs,floats
nofootinbib,tightenlines,superscriptaddress]{revtex4}

\usepackage{graphicx}
\usepackage{latexsym}

\newcommand{\comment}[1]{}

\newcommand{\beq}{\begin{equation}}
\newcommand{\eeq}{\end{equation}}
\newcommand{\bqa}{\begin{eqnarray}}
\newcommand{\eqa}{\end{eqnarray}}
\newcommand{\be}{\begin{equation}}
\newcommand{\ee}{\end{equation}}
\newcommand{\bea}{\begin{eqnarray}}
\newcommand{\eea}{\end{eqnarray}}
\newcommand{\nn}{\nonumber}
\newcommand{\0}{\over }

\newcommand{\6}{\partial }

\newcommand{\tr}{\,{\rm tr}\,}

\begin{document}

\title{Dynamics of
Quark-Gluon-Plasma Instabilities in %the
\\ Discretized Hard-Loop Approximation}

\preprint{TUW-05-09}
\preprint{HIP-2005-25/TH}
\preprint{BI-TP 2005/17}

\author{Anton Rebhan}
\affiliation{Institut f\"ur Theoretische Physik, Technische Universit\"at Wien,
        Wiedner Hauptstrasse 8-10, A-1040 Vienna, Austria}
\author{Paul Romatschke}
\affiliation{Fakult\"at f\"ur Physik, Universit\"at Bielefeld,
D-33501 Bielefeld, Germany}
\author{Michael Strickland}
\affiliation{Institut f\"ur Theoretische Physik, Technische Universit\"at Wien,
        Wiedner Hauptstrasse 8-10, A-1040 Vienna, Austria}
\affiliation{%Department of Physical Sciences and
Helsinki Institute of Physics \\
P.O. Box 64, FIN-00014 University of Helsinki, Finland}

\date{\today}

%\vspace{0.3cm}

\begin{abstract}
  Non-Abelian plasma instabilities have been proposed as a possible
  explanation for fast isotropization of the quark-gluon plasma
  produced in relativistic heavy-ion collisions. We study the
  real-time evolution of these instabilities in non-Abelian plasmas
  with a momentum-space anisotropy using a hard-loop effective theory
  that is discretized in the velocities of hard particles.  We extend
  our previous results on the evolution of the most unstable modes,
  which are constant in directions transverse to the direction of
  anisotropy, from gauge group SU(2) to SU(3).  We also present first
  full 3+1-dimensional simulation results based on velocity-discretized
  hard loops. In contrast to the
  effectively 1+1-dimensional transversely constant modes we find
  subexponential behavior at late times.
\end{abstract}

\maketitle
%\newpage

\section{Introduction}

If a quark-gluon plasma has been produced in the
current heavy-ion-collider experiments at RHIC,
the early onset of hydrodynamic behavior indicates
so fast thermalization, or at least isotropization,
that perturbative scattering processes alone
do not seem to be capable of explaining it
\cite{Baier:2000sb,Molnar:2001ux,Heinz:2004pj,Peshier:2005pp,Kovchegov:2005ss}.
A number of %more or less speculative 
nonperturbative 
scenarios have been put forward to account for the
seemingly inevitable intrinsic strong-coupling nature
of the quark-gluon plasma \cite{Shuryak:2004kh,Kovtun:2004de,Nastase:2005rp,Kharzeev:2005iz}.
However, a possible nonperturbative effect that does not
rely on specifically strong-coupling physics, and that has not
yet been fully taken into account in the available weak-coupling analyses, is
the important collective dynamics arising 
from non-Abelian plasma instabilities,
which are generalizations of the so-called Weibel or
filamentary instabilities in
ordinary electromagnetic plasmas \cite{Weibel:1959}.
Such instabilities are present already in {\em collisionless}
plasmas with any amount of momentum anisotropies
\cite{Romatschke:2003ms,Romatschke:2004jh}, and, indeed,
they have been found to play an important role in the fast isotropization
of electromagnetic plasmas \cite{Califano}.
Their non-Abelian versions 
have been proposed to be of relevance for the quark-gluon
plasma early on by Mr\'owczy\'nski and others
\cite{Heinz:1984,Stan:9397,Pokrovsky,Pavlenko:1991ih,
Mrowczynski:2000ed,Randrup:2003cw}, and specifically as explanation for
the recent puzzle of fast apparent thermalization by
Arnold, Lenaghan, Moore, and Yaffe \cite{Arnold:2003rq,Arnold:2004ti}.

At weak coupling, the asymptotic regime where perturbative QCD
becomes justifiable, there is a separation of scales between the
scale pertaining
to hard particles forming the nearly collisionless constituents of a plasma
and the scale which is down by one power of the QCD coupling $g$,
determining for example
electrical screening masses. 
The wave vectors and growth rates of Weibel instabilities
are of the same order, which is parametrically
much larger than perturbative collision rates or
even the parametrically larger color relaxation rate.
The degrees of freedom at the
scale $gp_{hard}$ are to leading order described by a non-local and non-linear
effective action called the hard-thermal-loop effective action
\cite{HTLeffact} in the case of thermal equilibrium, 
and this has a simple generalization to
plasmas with arbitrary momentum-space anisotropy \cite{Mrowczynski:2004kv}.

In the static limit, the anisotropic
hard-loop effective action has a potential which
is unbounded from below. Using a static approximation
and restricting to modes with momenta in the direction of
anisotropy, upon which
the hard-loop effective action becomes Gaussian and local, the evolution
of plasma instabilities has been studied numerically
by Arnold and Lenaghan in Ref.~\cite{Arnold:2004ih}. Based on these results,
these authors have conjectured that non-Abelian plasma instabilities
Abelianize in the regime of nonperturbatively large amplitudes, and
thus grow exponentially like their Abelian counterparts
as long as the assumptions of the hard-loop approximation hold.
At parametrically larger gauge-field amplitudes
$A\sim p_{hard}/g$, when they begin to have non-perturbatively
large effects on the hard particles' trajectories, these modes are expected to
eventually saturate by isotropizing the hard particle distribution.%
\footnote{This important aspect has recently been studied
in Ref.~\cite{Dumitru:2005gp} for non-Abelian plasma
instabilities. Our objective here is
to investigate the hard-loop dynamics preceding (in a weak-coupling
regime) the regime where backreaction needs to be included.}

The full hard-loop effective action, however, is nonlocal and
nonlinear. Using an auxiliary-field formalism to make it local
and discretizing with respect to the velocity space of the hard
particles, we have carried out a
full hard-loop simulation for an SU(2) plasma
in Ref.~\cite{Rebhan:2004ur} using initial conditions
that like those considered in
Ref.~\cite{Arnold:2004ih} are constant with respect to the spatial directions
transverse to the anisotropy axis. These configurations
evolve 1+1-dimensionally and contain the most unstable modes
for a given longitudinal momentum.
We found that there is subexponential behavior when
the instabilities reach nonlinear amplitudes, but exponential
growth roughly equal to that of the linear regime was restored
deeper into the nonlinear regime. However, the instabilities
of the full hard-loop effective theory
do not exhibit global Abelianization as in the
simplified model of Ref.~~\cite{Arnold:2004ih}, but only
in limited domains with extent comparable to the wavelength
of maximal growth.

In this paper we give a fuller account of these numerical
simulations and we present their extension to the gauge
group SU(3) of QCD. After introducing the local auxiliary-field
formulation of the hard-loop effective theory in Sect.~\ref{sect:HL},
we introduce our method to discretize hard-loop momenta
in Sect.~\ref{sect:DHL} and compare with continuum results
for the dispersion laws in Sect.~\ref{sect:DL}.
In Sect.~\ref{sect:1D}, we study the effectively 1+1-dimensional evolution of
initial configurations with random fluctuations in the
spatial direction of anisotropy, but which are constant with respect
to transverse directions. We analyse discretization effects and
the effect of upgrading the gauge group SU(2) to SU(3).
In Sect.~\ref{sect:3D} we finally consider initial conditions
that are non-constant 
in the transverse direction and thus require full 3+1-dimensional simulations.
While at the beginning of the nonlinear regime we obtain
results similar to the effectively 1+1-dimensional
simulations, we observe subexponential behavior deeper into
the nonlinear regime, in qualitative agreement with a most recent 
3+1-dimensional study
by Arnold et al.~\cite{Arnold:2005vb} using a different discretization method.
Sect.~\ref{sect:Concl} contains our conclusions.

\section{Hard-loop effective field equations for
anisotropic plasmas}\label{sect:HL}

At weak gauge coupling $g$, there is a separation of
scales in hard momenta $|\mathbf p|=p^0$ of (ultrarelativistic)
plasma constituents, and soft momenta $\sim g|\mathbf p|$
pertaining to collective dynamics such as Debye screening, finite
plasma frequency, and, in anisotropic plasmas, magnetic instabilities.
The effective field theory for the soft modes that is
generated by integrating out the hard plasma modes
at one-loop order and in the approximation that
the amplitudes of the soft gauge fields obey $A_\mu \ll |\mathbf p|/g$
is that of gauge-covariant collisionless Boltzmann-Vlasov equations
\cite{HTLreviews}.
In equilibrium, the corresponding (nonlocal) effective action is
the so-called hard-thermal-loop effective action 
\cite{HTLeffact%Taylor:ia,Braaten:1991gm,Frenkel:ts
} which has a simple generalization
to plasmas with anisotropic momentum distributions \cite{Mrowczynski:2004kv}.
At the expense of introducing a continuous set of auxiliary fields,
the hard-loop effective field equations can be made local
in space and time. These field equations then involve 
an induced current of the
form
\cite{Blaizot:1993be,Mrowczynski:2000ed}
\begin{equation}\label{Jind0}
j^\mu[A] = -g^2
\int {d^3p\over(2\pi)^3} 
{1\over2|\mathbf p|} \,p^\mu\, {\partial f(\mathbf p) \over \partial p^\beta}
W^\beta(x;\mathbf v)
\end{equation}
with a set of auxiliary fields $W_\beta(x;\mathbf v)$ for
each spatial unit vector appearing in the velocity 
$v^\mu=p^\mu/|\mathbf p|=(1,\mathbf v)$ of
a hard (ultrarelativistic) particle with momentum $p^\mu$. The $W$'s satisfy
\begin{equation}\label{Weq}
[v\cdot D(A)]W_\beta(x;\mathbf v)
= F_{\beta\gamma}(A) v^\gamma \;.
\end{equation}
with $D_\mu=\partial_\mu-ig[A_\mu,\cdot]$.
%We shall eventually use temporal gauge, so that $A^\mu=(0,\mathbf A)$.

The equations are closed by the non-Abelian Maxwell equations
\beq\label{Feq}
D_\mu(A) F^{\mu\nu}=j^\nu,
\eeq
where $\nu=0$ is the Gauss law constraint.\footnote{Our metric convention
is $(+---)$.}

In the following we shall consider the special case of
cylindrically symmetric anisotropies in momentum space, 
so that there is only one direction of anisotropy, which in
heavy-ion collision would be given by the collision axis.
We can then
parametrize $f(\mathbf p)= \tilde f(|\mathbf p|,p^z)$
and write
\bea
{\partial f(\mathbf p) \over \partial p^\beta}
&=&
{\partial \tilde f(|\mathbf p|,p^z) \over \partial |\mathbf p|}
{p^b \delta_{b\beta}\0|\mathbf p|}
+{\6 \tilde f(|\mathbf p|,p^z) \over \6 p^z}\delta_{z\beta}\nonumber\\
&\equiv&
-\tilde f_1(|\mathbf p|,p^z)
{p^b \delta_{b\beta}\0|\mathbf p|}
-\tilde f_2(|\mathbf p|,p^z)\delta_{z\beta}.
\eea

Eq.~(\ref{Weq}) allows us to impose $p^\beta W_\beta=0$, which leads to
\be\label{Jind2}
j^\mu(x)={1\02}g^2
\int {d^3p\over(2\pi)^3} 
v^\mu\, [\tilde f_1 W^0(x;v)+\tilde f_2W^z(x;v)].
\ee

Notice that in the isotropic case one has $f_2=0$, and only $W^0$
appears, whose equation of motion (\ref{Weq}) involves only electric
fields. In the anisotropic case, however, $W^z$ enters, whose
equation of motion contains the $z$ component of the 
Lorentz force.

The induced current (\ref{Jind0}) is covariantly conserved
for arbitrary momentum-space distribution functions $f$ as
can be verified by partial
integration with respect to $p^\beta$
\cite{Mrowczynski:2004kv}. In the case of a parity-invariant
function $f$, $f(\mathbf p)=f(-\mathbf p)$, 
to which we shall restrict ourselves in the following,
no partial integration
is needed and, moreover,
the two terms in the current (\ref{Jind2}) are covariantly 
conserved individually:
Using (\ref{Weq})
we have
\be\label{Dj}
D\cdot j \propto
\int {d^3p\over(2\pi)^3} (\tilde f_1 F^{0\gamma}v_\gamma+
\tilde f_2 F^{z\gamma}v_\gamma)=0
\ee
where parity invariance of $f_1$ makes the $\mathbf v$ in the
first term integrate to zero. In the second term $f_2$ is odd with
respect to the $z$-direction, so that only
the term involving $F^{zz}v_z$ could contribute, but $F^{zz}\equiv 0$.

The dynamical system described by Eqs.~(\ref{Jind0})--%
(\ref{Feq}) has constant total energy of the form
\be\label{Etot}
\mathcal E=
\int d^3x \tr (\mathbf E^2 + \mathbf B^2)|_t + 
\int_{t_0}^t\! dt' \int d^3x\, 2 \tr \mathbf j_{t'}\cdot \mathbf E_{t'}.
\ee
The part containing the induced current
%can be written as
involves
\bea
\tr \mathbf j\cdot \mathbf E
&=&{g^2\04}\6_\mu \int {d^3p\over(2\pi)^3} \tilde f_1 v^\mu W_0^2\nonumber\\
&&+{g^2\02}\int {d^3p\over(2\pi)^3} \tilde f_2 W^z (v\cdot D)W_0,
\eea
which shows that in the isotropic case ($\tilde f_2=0$) there is
a local, positive definite energy contribution from the plasma
\cite{Blaizot:1994am}.
However, in the general anisotropic case, positivity is lost.
This reflects
the possibility of plasma instabilities, where
energy may be extracted from hard particles and deposited into
the soft collective fields without bound
as long as the hard-loop approximation $A_\mu \ll |\mathbf p|/g$
remains valid.
%(which requires that $A_\mu \ll |\mathbf p|/g$).
%Actually, it
%is still possible to write the hard-loop energy 
%as a local quantity in terms of $W_\mu$'s and
%derivatives of $W_\mu$'s with respect to $\mathbf v$ \cite{Blaizot:1994am},
%but for the following applications it will be more convenient
%to integrate over time rather than consider differentiations with
%respect to $\mathbf v$.

%\section{Discretization}

\section{Discretized hard loops}\label{sect:DHL}

The structure of Eqs.~(\ref{Jind0})--(\ref{Feq}) is such that only 
$W_0$ and $W_z$ participate nontrivially in the dynamical evolution.
A closed set of gauge-covariant
equations is thus obtained when the hard-loop integral in (\ref{Jind2})
is discretized with respect to directions $\mathbf v$
\bea\label{Jindd}
j^\mu(x)&=&g^2
\int {p^2 dp\over(2\pi)^2} {1\0\mathcal N} \sum_{\bf v} 
v^\mu\, [\tilde f_1 W^0_{\mathbf v}(x)+
\tilde f_2 W_{\mathbf v}^z(x)]\nonumber\\
&&\equiv {1\0\mathcal N} \sum_{\bf v} v^\mu[a_{\mathbf v} W^0_{\mathbf v}(x)+
b_{\mathbf v} W^z_{\mathbf v}(x)],
\eea
where the $\mathcal N$ unit vectors $\mathbf v$ define a partition
of the unit sphere in patches of equal area.
The coefficients $a_{\mathbf v}$, $b_{\mathbf v}$ are constants
defined by
\bea\label{abv}
a_{\mathbf v}&=&-g^2\int_0^\infty {p^2 dp\over(2\pi)^2} 
f_1(|\mathbf p|,|\mathbf p|v^z), \\
b_{\mathbf v}&=&-g^2\int_0^\infty {p^2 dp\over(2\pi)^2} 
f_2(|\mathbf p|,|\mathbf p|v^z). \eea
Isotropic distribution functions $f$ are characterized by
$a_{\mathbf v}$'s which are independent of $\mathbf v$, and
vanishing $b_{\mathbf v}$'s.

The special choice of Ref.~\cite{Romatschke:2003ms,Romatschke:2004jh}
for an anisotropic distribution function, 
\be\label{faniso}
\tilde f(|\mathbf p|,p^z)=N(\xi) f_{iso}(\mathbf p^2+\xi p_z^2),
\ee
gives
\be\label{avbvxi}
a_{\mathbf v}={N(\xi)m_D^2\0(1+\xi v_z^2)^2},\qquad
b_{\mathbf v}=\xi v_z a_{\mathbf v},
\ee
where $m_D$ is the Debye mass of the isotropic case $\xi=0$.
If one requires that the number density of hard particles
is the same for different values $\xi$, the normalization
factor here is $N(\xi)=\sqrt{1+\xi}$. In the following
we shall often use the abbreviation 
\be\label{m2norm}
m^2\equiv N(\xi)m_D^2,
\ee
leaving open the possibility for different normalizations.\footnote{%
Note that Ref.~\cite{Romatschke:2003ms} has $N(\xi)=1$.}

Discretization of the directions $\mathbf v$ can
spoil the reflection properties of $f_{1,2}$ and thus
automatic covariant conservation of $j$ by virtue of
Eq.~(\ref{Dj}). Covariant current conservation
is however guaranteed when
$a_{\mathbf v}$, $b_{\mathbf v}$ satisfy
\be\label{avsum}
\sum_{\mathbf v} a_{\mathbf v} \mathbf v = 0,\qquad
\sum_{\mathbf v} b_{\mathbf v}  = 0,\qquad
\sum_{\mathbf v} b_{\mathbf v} \mathbf v_\perp  = 0,
\ee
where $\mathbf v_\perp = \mathbf v - v^z \mathbf e_z$,
and so we shall restrict ourselves to discretizations
of this kind.

A suitable discretization of the sphere is given by a set
of unit vectors $\mathbf v$ following from
regular spacing in $z$ and $\varphi$ according to
\be\label{disco}
z_i=-1+(2i-1)/N_z,\quad i=1\ldots N_z,\qquad
\varphi_j=2\pi j/N_\varphi,\quad j=1\ldots N_\varphi.
\ee
The resulting ``disco balls'' are such that they are covered with
$\mathcal N=N_z\times N_\varphi$ (curved) tiles of equal area.
For low values of $\mathcal N$,
we shall also consider discretized $\mathbf v$'s pointing
to the centers of the faces of one of the
regular polyhedra with $\mathcal N=6,8,12$ or 20.

Given any set of $\mathbf v$, $a_{\mathbf v}$, $b_{\mathbf v}$,
satisfying (\ref{avsum}),
one has then to solve (\ref{Jindd}) together with
the Yang-Mills field equations and
\begin{equation}\label{Weqd}
[v\cdot D(A)]W^{\mathbf v}_\beta(x)
= F_{\beta\gamma}(A) v^\gamma, \qquad \beta=0,3.
\end{equation}

In temporal axial gauge $A^0=0$, the dynamical variables are
$A_i$, $E_i=-\dot A_i$, $W^{\mathbf v}_0$ and $W^{\mathbf v}_z$.
Eq.~(\ref{Weqd}) becomes
\be
\6_t W^{\mathbf v}_\beta(x)=-\mathbf v\cdot \mathbf D\; W^{\mathbf v}_\beta
+F_{\beta\gamma}(A) v^\gamma.
\ee
The gauge choice $A^0=0$ is in fact merely a convenience for
the following numerical simulations. All equations are gauge covariant,
the current is covariantly conserved, and we shall also
consider exclusively gauge invariant observables in our
numerical simulations.

Through Eq.~(\ref{Jindd}) only a linear combination of $W^0_{\mathbf v}$
and $W^z_{\mathbf v}$ participates actively in the dynamical evolution.
So in addition to the gauge fields
we need to consider only the $\mathcal N$ auxiliary fields
\be
\mathcal W_{\mathbf v}(x)=a_{\mathbf v} W^0_{\mathbf v}(x)+
b_{\mathbf v} W^z_{\mathbf v}(x).
\ee

The full hard-loop dynamics is then approximated
by the following set of matrix-valued equations,
\bea\label{DHLW}
&&[v\cdot D(A)]\mathcal W_{\mathbf v}=(a_{\mathbf v} F^{0\mu}+b_{\mathbf v}
F^{z\mu})v_\mu\\
\label{DHLF}
&&D_\mu(A) F^{\mu\nu}=j^\nu={1\0\mathcal N}\sum\limits_{\mathbf v} v^\nu 
\mathcal W_{\mathbf v},
\eea
which can be systematically improved by increasing $\mathcal N$.

The Gauss law constraint is explicitly
\be
D_i(A)F^{i0}=-D_i(A)\dot A^i=j^0=
{1\0\mathcal N} \sum_{\bf v} \mathcal W_{\mathbf v}.
\ee

In the case of an isotropic plasma, a similar discretization of the
hard-loop dynamics which uses
a discrete set of vectors $\mathbf v$
has been considered before by Rajantie and Hindmarsh
\cite{Rajantie:1999mp}.
A different possibility for discretization that
has been employed previously is 
a decomposition of the auxiliary fields $W^\mu(x;v)$
into spherical harmonics
\cite{Bodeker:1999gx,Rajantie:1999mp} and truncating
at a maximal angular momentum $l_{max}$.
This method has now also been used in the most recent numerical
study of non-Abelian plasma instabilities
by Arnold et al. \cite{Arnold:2005vb}.
Discretizing $\mathbf v$ is slightly simpler, but also more
flexible in that it allows one to e.g.\ improve approximations in highly
anisotropic cases with cylindrical symmetry by selectively
increasing the resolution in the $z$ direction more than in the $\varphi$
direction.

%\subsection{1+1-dimensional dynamics}

%So far all formulas have been fully 3+1 dimensional.
When the momentum space distribution of hard modes has an oblate form,
$\xi>0$ in Eq.~(\ref{faniso}),
the momenta of the unstable collective modes form prolate regions
with $|k_z|>0$ \cite{Romatschke:2003ms,Arnold:2003rq}.
For a given $k_z$, the growth rate is largest for $k_\perp=0$,
i.e.\ the constant modes with respect to the directions
transverse to the direction of momentum space anisotropy.
By considering initial conditions which are constant in the
transverse direction, we can study the complete dynamics
of these particular unstable modes in a 1+1-dimensional
setting where all fields are dimensionally reduced,
$(x)\to(t,z)$. Then only $A_z(t,z)$ plays the role
of a gauge field, and $A_{x,y}(t,z)$ behave as adjoint matter.

Using temporal axial gauge, %$A^0=0$,
the equations of motion for the dynamical fields can then be simplified
according to
\bqa\label{1+1eoms}
\partial_{t}^{2} \, A_x &=& D_z^2 A_x-g^2 [A_y,[A_y,A_x]]+j^x\nonumber\\
\partial_{t}^{2} \, A_y &=& D_z^2 A_y-g^2 [A_x,[A_x,A_y]]+j^y\nonumber\\
\partial_t^2 \, A_z &=&-i g [A_x,D_z A_x]-i g [A_y,D_z A_y]+j^z\nonumber\\
(\partial_t +  {\mathbf v}\cdot{\mathbf D} )
%\{\underbrace{a_{\mathbf v}W_{\mathbf v}^0 
%+b_{\mathbf v}W_{\mathbf v}^z}_{\mathcal W_{\mathbf v}}\}
\mathcal W_{\mathbf v}
&=& a_{\mathbf v} ( -{\mathbf v}\cdot{\partial_t {\mathbf A}} )
+b_{\mathbf v} [
-\partial_t A_z+
%\underbrace{D_z {\mathbf v}\cdot{\mathbf A} 
%-({\mathbf v}\cdot{\mathbf \nabla}) A_z}_{D_z(v_x A_x+v_y A_y)}
D_z(v_x A_x+v_y A_y)
 ],
%W_{\mathbf v}^0-{\mathbf v}\cdot{\partial_t {\mathbf A}} \nonumber \\
%\partial_t \, W_{\mathbf v}^z &=&-{\mathbf v}\cdot{\mathbf D}\, W_{\mathbf v}^z
%-\partial_t A_z+D_z {\mathbf v}\cdot{\mathbf A}
%%-({\mathbf v}\cdot{\mathbf \partial}) A_z,
\eqa
\comment{mpt}
where ${\mathbf D}={\nabla}+i g [{\mathbf A},\cdot]$,
$\nabla=(0,0,\partial_z)$,
and ${\mathbf j}
={1\0\mathcal N}\sum\limits_{\mathbf v} \mathbf v \mathcal W_{\mathbf v}$.

%The Gauss law constraint in temporal axial gauge is
%\be
%-\mathbf D \cdot \dot \mathbf A = {1\0\mathcal N}\sum_{\mathbf v} 
%\mathcal W_{\mathbf v}.
%\ee
%\comment{mpt}

In Sect.~\ref{sect:3D} we shall also consider initial conditions
which are not constant with respect to the transverse directions.
This will then require the full 3+1-dimensional content of
Eqs.~(\ref{DHLW}) and (\ref{DHLF}).

\section{Comparison of dispersion laws}\label{sect:DL}

Before turning to the numerical evaluation of the above
equations by also discretizing space and time, we shall
study the dispersion laws of linearized modes, and
the consequences of replacing
the continuous set
of fields $W^\beta(x,\mathbf v)$ by a finite number $W^\beta_{\mathbf v}(x)$.
Because we shall be interested mainly in the unstable modes,
we shall concentrate on wave-vectors 
parallel to the anisotropy axis, which includes the most
unstable modes. With
$k^i=k\delta^i_z$, there are then only two possible modes
for gauge fields, transverse and longitudinal ones
\cite{Romatschke:2003ms,Romatschke:2004jh}.

\subsection{Continuum results}

For anisotropic distribution functions (\ref{faniso}),
analytical results for the gluonic self-energies
have been obtained in Ref.~\cite{Romatschke:2003ms,Romatschke:2004jh},
which we recapitulate for the special case of longitudinal
wave vectors.

\subsubsection{Transverse polarizations}

With $\eta=\omega/k$, the transverse gluon self-energy reads
\cite{Romatschke:2004jh}
\bqa\label{PiTaniso}
\Pi_t&=&\frac{m^2}{4 \sqrt{\xi} (1+\xi \eta^2)^2} %
\left[\left(1+\eta^2+\xi (-1+(6+\xi) \eta^2-(1-\xi) \eta^4)\right) %
\arctan \sqrt{\xi}\right.\nonumber \\
&& \hspace{4cm} \left.+ \sqrt{\xi}\, (\eta^2-1) 
\left(1+\xi \eta^2-(1+\xi) \eta \ln{\frac{%
\eta+1+i \epsilon}{\eta-1+i \epsilon}}\right)\right] \, , \nonumber \\
\eqa
where we recall that $m^2=N(\xi)m_D^2(\xi\!=\!0)$.

The limit $k\to0$ or $\eta\to\infty$ determines
the transverse plasma frequency as\footnote{%
For $-1<\xi<0$ one has to replace ${1\0\sqrt{\xi}}\arctan{\sqrt\xi}$
by ${1\0\sqrt{-\xi}}{\rm artanh}{\sqrt{-\xi}}$.}
\be\label{ompltconti}
\omega^2_{pl,t}={1\04\xi}\left(1+{\xi-1\0\sqrt{\xi}}\arctan{\sqrt\xi}\right)
m^2.
\ee
%whose small-$\xi$ expansion reads
%\be\label{ompltcont}
%\omega^2_{pl,t}/m^2=
%\underbrace{1\03}_{p_1}-\underbrace{2\015}_{p_2}\xi+\underbrace{3\035}_{p_3}\xi^2+O(\xi^3).
%\ee
%(The coefficients $p_k$ will be compared with the discretized
%results in Table \ref{tableA}). 

%Propagating modes with small $k$ are characterized by
%\be\label{omkcont}
%\omega^2_t(k)=\omega^2_{pl,t}+\Bigl[\underbrace{5\06}_{q_1}
%+\underbrace{19\0175}_{q_2}\xi+O(\xi^2)\Bigr]k^2+O(k^4).
%\ee

For large momenta, the transverse excitations tend to the
asymptotic mass given by
\be\label{masconti}
m^2_\infty/m^2={1\02\sqrt{\xi}}\arctan{\sqrt\xi}.
\ee
%whose small-$\xi$ expansion reads
%\be\label{mascont}
%m^2_\infty/m^2=\underbrace{1\02}_{r_1}-\underbrace{1\06}_{r_2}\xi
%+\underbrace{1\010}_{r_3}\xi^2+O(\xi^3)
%\ee

The static limit on the other hand 
gives a mass squared which is determined by the relation
\be\label{mmvsmpl}
m_m^2\equiv -\mu^2=-\xi \omega^2_{pl,t}\;.
%-{1\04}\left(1+{\xi-1\0\sqrt{\xi}}\arctan{\xi}\right)
\ee
For negative $\xi$, this indicates screening of magnetostatic fields,
while for positive $\xi$ there is an instability for all momenta
$k<\mu$.

\subsubsection{Longitudinal polarizations}

In the longitudinal sector we have \cite{Romatschke:2004jh}
\bqa
\Pi_\ell&=&- \frac{\eta^2 m^2}{2 \sqrt{\xi} (1+\xi \eta^2)^2} %
\left[(1+\xi)(1-\xi \eta^2)\arctan{\sqrt{\xi}} \right. \nonumber \\
&& \hspace{4cm} \left. + \sqrt{\xi} \left(%
(1+\xi \eta^2)-(1+\xi) \eta \ln{\frac{\eta+1+i \epsilon}{\eta-1+i \epsilon}}
\right)\right] \, .
\eqa
The plasma frequency in the longitudinal sector is generally different
from the one in the transverse sector when $\xi\not=0$ and reads
\be\label{ompllconti}
\omega^2_{pl,\ell}={1\02\xi} \left({1+\xi \0 \sqrt\xi} \arctan{\sqrt\xi} - 1
\right)m^2\;.
\ee
%whose small-$\xi$ expansion is
%\be\label{ompllcont}
%\omega^2_{pl,\ell}/m^2=
%\underbrace{1\03}_{s_1}-\underbrace{1\015}_{s_2}\xi
%+\underbrace{1\035}_{s_3}\xi^2+O(\xi^3).
%\ee

The static limit of $\Pi_\ell/\eta^2$ gives the electric screening mass
squared
\be
m^2_{el}={1\02}\left({1+\xi \0 \sqrt\xi} \arctan{\sqrt\xi}+1
\right)m^2
=m^2+\xi \omega^2_{pl,\ell}\;.
\ee
This parameter does not change sign for all $-1\le\xi<\infty$, so
in the present case of wave vectors parallel to the anisotropy direction
all instabilities are purely transverse.

\subsection{$\mathbf v$-discretized hard loops}

After discretizing the space of velocities and
approximating the fields $W^\beta(x,\mathbf v)$
by a finite set of field $W^\beta_{\mathbf v}(x)$, we search
for solutions of Eqs.~(\ref{1+1eoms}) where all fields point in
a fixed color direction and which correspond to a single Fourier
mode in $z$-direction:
\be
A^i(t,z)=e^{\gamma t}e^{ikz} \epsilon^i,\quad
W_{\mathbf v}^{0,z}(t,z)=e^{\gamma t}e^{ikz} w_{\mathbf v}^{0,z}.
\ee
Here $k$ will usually be a real number, while $\gamma$
can be either real or imaginary, corresponding to
exponential or oscillatory behavior, respectively.
The dynamics is then effectively Abelian and linearized, and
Eqs.~(\ref{1+1eoms}) reduce to
\bea\label{fteoms}
\gamma^2\epsilon^i+k^2\epsilon^i_\perp&=&
{1\0\mathcal N}\sum\limits_{\mathbf v} v^i
(a_{\mathbf v}w_{\mathbf v}^0 
+b_{\mathbf v}w_{\mathbf v}^z)\\
(\gamma+ikv_z)w_{\mathbf v}^0&=&-\gamma v^i \epsilon^i\\
(\gamma+ikv_z)w_{\mathbf v}^z&=&-\gamma \epsilon^z
%+ik(v^x\epsilon^x+v^y\epsilon^y)
+ik v^i \epsilon^i_\perp
\eea
with $\epsilon^i_\perp=(\epsilon^x,\epsilon^y,0)$.
The Gauss law constraint reads
\be\label{ftglc}
-ik\gamma\epsilon^z={1\0\mathcal N}\sum\limits_{\mathbf v}
(a_{\mathbf v}w_{\mathbf v}^0 +b_{\mathbf v}w_{\mathbf v}^z).
\ee

Next we specialize to the anisotropic distribution
function %of Ref.~\cite{Romatschke:2003ms,Romatschke:2004jh}
(\ref{faniso}),
%for which there are analytical results for the dispersion laws, and
for which the coefficients $a_{\mathbf v}$, $b_{\mathbf v}$
are given by (\ref{avbvxi}), and compare with the above
continuum results.
With discretized $\mathbf v$'s we obtain algebraic equations
that can be solved in closed form for the regular polyhedra 
with $\mathcal N=6,8,12,20$. Taking the $\mathbf v$'s to
point to the centers of the faces of these polyhedra, we
consider
two different orientations for each polyhedron: one where the discrete
rotational symmetry around the $z$-axis is maximal, and one where
this symmetry is only $Z_3$. The two choices sample in particular different
sets of $z$-components of hard momenta (in Table \ref{tableA}
in Appendix \ref{appdhlcomp}
the number of distinct $z$-components is indicated by $N_z$).
We shall also work out the dispersion laws in
a few cases of larger $\mathcal N$ 
using Eqs.~(\ref{disco}).

\subsubsection{Transverse polarizations}

Equations (\ref{fteoms}) give algebraic relations between $\gamma$
and $k$, while the Gauss law constraint is automatically
fulfilled for reflection-symmetric distribution functions,
which lead to the restrictions (\ref{avsum}).

For the cube ($\mathcal N=6$) 
and $Z_4$ symmetry about the $z$-axis, the solution
is simply
\be\label{disp6a}
\gamma^2=-\omega^2=-(k^2+m^2/3),
\ee
corresponding to a propagating mode with momentum-independent mass equal
to the isotropic plasma frequency. In this case, the
$\xi$-dependence drops out completely (except for a possible
normalization of $m$ according to (\ref{m2norm})). The result (\ref{disp6a})
is in fact identical to the one obtained
by discretization through spherical harmonics and
truncating at angular momentum $l_{max}=1$ \cite{Bodeker:1999gx}.

\begin{figure}
\includegraphics[width=2.5in]{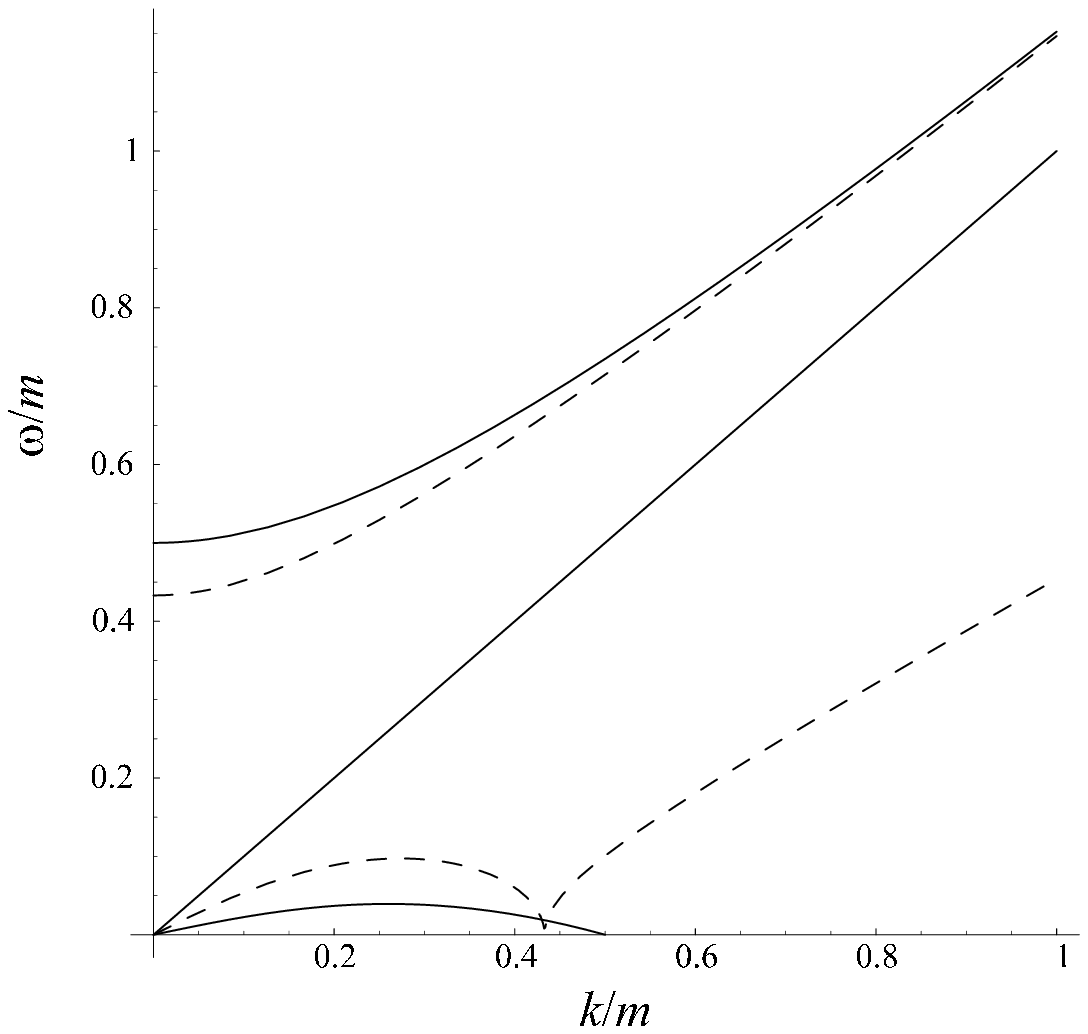}
\includegraphics[width=2.5in]{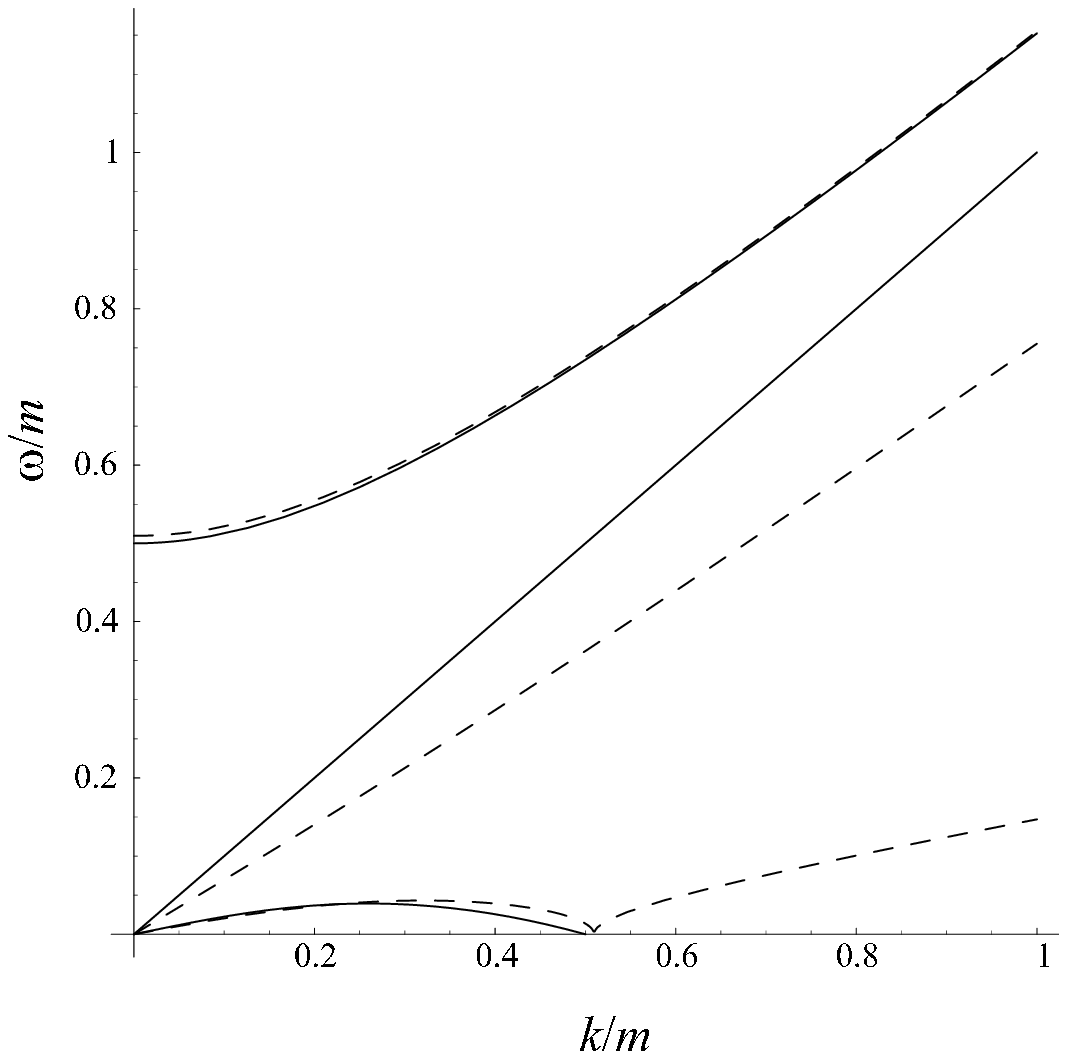}
\centerline{\hfil (a) $\mathcal N=8$ \hfil\hfil (b) $\mathcal N=20$ \hfil}
\caption{The transverse dispersion laws $\omega(k)$ for anisotropy
parameter $\xi=1$
in the continuum (full lines) compared with those
of $\mathbf v$-discretized hard loops for the
octahedron ($\mathcal N=8$, $Z_4$) and 
the icosahedron ($\mathcal N=20$, $Z_5$) (dashed lines). The unstable modes
appear at small space-like momenta where the plot is
in terms of $\gamma(k)$ instead of $\omega(k)$.
The additional space-like modes are approximations of the
Landau cut by simple poles whose number increases with $\mathcal N$.
\label{figdisptr}}
\end{figure}

For $\mathcal N=6$ and $Z_3$ symmetry about the $z$-axis, the
solution is more complicated, and happens to coincide
with the $\mathcal N=8$, $Z_4$ case. These two cases give
the biquadratic equation
\be
\gamma^2+k^2=\underbrace{m^2\0(1+\xi/3)^2}_a
{-\gamma^2+\xi k^2/3\03\gamma^2+k^2}\,.
\ee
This equation obviously has solutions for real $\gamma$
(i.e.\ instabilities) only when $\xi>0$, as is the case
for the continuum solutions.
There are two branches of solutions given by
\be\label{N8Z4eq}
\gamma^2\equiv-\omega^2
=-{1\06}\left[
a+4k^2\pm\sqrt{a^2+(8+4\xi)ak^2+4k^4}\right]
\ee
The upper sign corresponds to propagating transverse plasmons.
%with
%coefficients $p_k,q_k,r_k$ as given in Table \ref{tableA}.
As can be seen from
Table \ref{tableA} in Appendix \ref{appdhlcomp}, 
there is considerable improvement
compared to the simple dispersion law (\ref{disp6a}) of the
$\mathcal N=6$, $Z_4$ case --- the asymptotic mass and its small-$\xi$
correction are even exact. 
The lower sign in (\ref{N8Z4eq}) contains the instability for $\xi>0$,
and oscillatory behavior for $k>\mu$.
Fig.~\ref{figdisptr}a
shows a plot of the solutions of (\ref{N8Z4eq}) and compares it
with the continuum result following from (\ref{PiTaniso}).

The $\mathcal N=8$, $Z_3$ case and the $\mathcal N=12$, $Z_5$ case also lead to
biquadratic equations, each with somewhat different
coefficients, and the small-$\xi$ behavior can be
read from Table \ref{tableA}.

The $\mathcal N=12$, $Z_3$ case and
the icosahedron, $\mathcal N=20$, on the other hand, lead to bicubic equations,
with $\mathcal N=12$, $Z_3$ and $\mathcal N=20$, $Z_5$ sharing
the same equation\footnote{As is well known, 
the regular polyhedra $\mathcal N=6,8$ and $\mathcal N=12,20$ form
dual pairs, however this does not seem to explain the
coincidence of the dispersion laws of $\mathcal N=6$, $Z_3$ with
 $\mathcal N=8$, $Z_4$ and of $\mathcal N=12$, $Z_3$ with
$\mathcal N=20$, $Z_5$, since duality would map orientations
with rotational symmetry $Z_3$ to such with $Z_3$.}, namely
\be\label{N20Z5eq}
\gamma^2+k^2
=-{15m^2} { (45\gamma^4-k^4\xi)(45+42\xi+13\xi^2)+
3\gamma^2 k^2 (315+255\xi+93\xi^2-7\xi^3) 
\0 (45+30\xi+\xi^2)^2 (45\gamma^4 + 30\gamma^2 k^2+k^4) }.
\ee
The case $\mathcal N=20$, $Z_3$ gives a similar equation but with different
coefficients.
The solutions to these equations can be given in closed form, but are
too unwieldy to be reproduced here.
There are now three distinct solutions one of which corresponds to
%One corresponds to 
propagating transverse plasmons.
% characterized by the
%coefficients $p_k,q_k,r_k$ listed in Table \ref{tableA}
%of Appendix \ref{appdhlcomp}.

Except for the cube with $Z_4$ symmetry, Eq.~(\ref{disp6a}),
there is always exactly one solution which contains an instability
%(positive $\gamma^2$) 
for $\xi>0$ and $k<\mu$. 
A significant difference to the continuum result is however the
absence of Landau damping. Instead, there are additional
modes for spacelike momenta, and their number increases with $\mathcal N$,
approximating the logarithmic cut by more and more
poles at spacelike momenta. 
This is analogous to what has been observed in Ref.~\cite{Bodeker:1999gx}
for discretizations using spherical harmonics.
However, in the anisotropic case, one of these spacelike modes becomes
unstable for $\xi>0$. In the continuum limit, the unstable part ($k\le \mu$)
remains on the physical sheet, whereas its stable, oscillatory
part moves to the unphysical sheet, cf.~Ref.~\cite{Romatschke:2004jh}.

In 
Fig.~\ref{figdisptr}b we show the transverse dispersion laws for $\xi=1$
following from (\ref{N20Z5eq}), which has two spacelike branches.
The comparison with the
continuum case shows that both the timelike propagating modes and the
instabilities are quite well reproduced quantitatively.

\begin{figure}

\includegraphics[width=\textwidth]{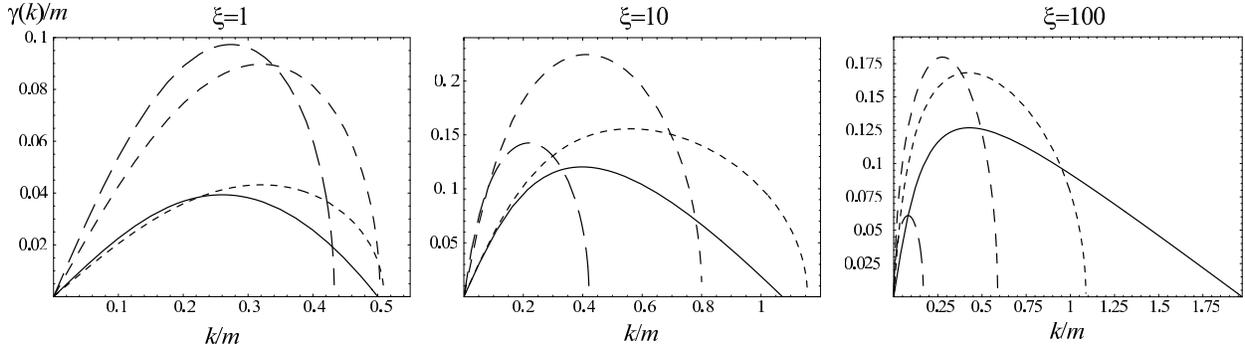}
\caption{The growth rate $\gamma(k)$ of unstable modes for the
polyhedral approximations $\mathcal N=8,12,20$
and maximal $Z_\varphi$ (dashed lines with smaller
and smaller dashes) in comparison with the continuum result (black lines)
for $\xi=1,10,$ and 100.
\label{figgammas}}
\end{figure}

Our main interest will clearly be the plasma instabilities. Since
all the actual and additional stable modes are just oscillatory,
we do not expect them to play an important role in the evolution of
the instabilities. 
For us the most important aspect of the dispersion laws
will be the magnitude of $\gamma(k)$ for the unstable modes $k\le \mu$.
The continuum result for $\gamma(k)$ has the property that it vanishes
both at $k=0$ and at $k=\mu$, in contrast with the toy model
studied in Ref.~\cite{Arnold:2004ih}. This feature is correctly
reproduced in our discretization of the hard-loop dynamics of
anisotropic plasmas. In Figs.~\ref{figgammas} we compare $\gamma(k)$
of the various polyhedra (with maximal $Z_\varphi$) with the exact continuum 
result for some values of $\xi$.
From this we expect that all the polyhedral approximations should
be qualitatively correct, with $\mathcal N=20$ being a quantitatively
good approximation for $\xi\lesssim10$.

\begin{figure}
\includegraphics[width=3in]{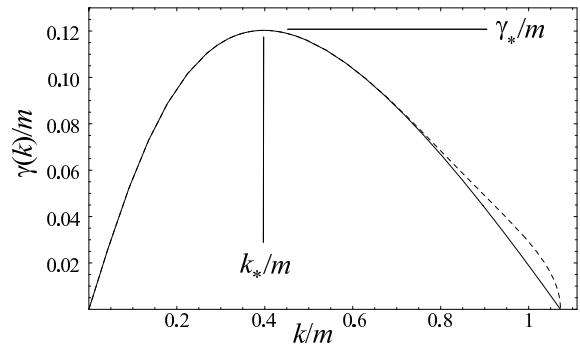}
\caption{The growth rate $\gamma(k)$ of unstable modes for the discretized
hard loops with
$\mathcal N=100$ (dashed line)
in comparison with the continuum result (solid line)
for $\xi=10$. The precise values of wave number $k_*$ and growth
rate $\gamma_*$ corresponding to the mode of maximal growth can be
read from Table~\ref{tablem}.
\label{figgamma}}
\end{figure}

For higher $\mathcal N$ we use the ``disco-ball'' discretization (\ref{disco}).
Fig.~\ref{figgamma} compares the growth rates $\gamma(k)$
for full and discretized hard loops with
$\mathcal N=\mathcal N_z \times\mathcal N_\varphi=20\times5=100$
and $\xi=10$,
which shows that the latter give very accurate approximations
for $\mathcal N\gtrsim 100$.
Also the dispersion laws of propagating stable modes are
accurately reproduced. For example, with $\xi=10$, $\mathcal N=100$,
the continuum result for the asymptotic mass (\ref{mascont}) of the transverse
modes with $|\mathbf k| \gg \mu$
is reproduced with an error of only 0.017\% 

In the following numerical simulations, we shall employ the
disco-ball discretization with $\mathcal N\ge 100$, 
choosing $N_z \gg N_\varphi$
to take into account that the distribution function varies
more rapidly in the $z$ than in the $\varphi$-direction.
When considering smaller $\mathcal N$,
we shall instead take regular polyhedra, which
as discussed further in Appendix \ref{appdhlcomp},
seem to be superior approximations for $\mathcal N\le 20$. 
Before turning to these numerical simulations,
we complete this section by briefly considering also
the (stable) longitudinal modes.

\subsubsection{Longitudinal polarizations}

For longitudinal polarizations ($\epsilon^z\not=0$) the Gauss law
constraint (\ref{ftglc}) is nontrivial, but it leads to the
same equation as the linearized field equation (\ref{fteoms}).
For the regular polyhedra, the resulting algebraic relations
between $\gamma$ and $k$ can be solved in closed form, and
one finds that as expected there are only stable solutions $-\gamma^2
=\omega^2\ge0$.

For $\mathcal N=6$ and $Z_4$ symmetry about the $z$-axis, the solution
is 
\be\label{displ6a}
\omega^2={m^2\03(1+\xi)}+k^2,
\ee
corresponding to a propagating mode with momentum-independent mass 
which at $\xi=0$ equals the correct plasma frequency. In the
static limit, the Debye mass is off by a factor $1/\sqrt{3}$.

The cases $\mathcal N=6$ with $Z_3$ symmetry about the $z$-axis
and $\mathcal N=8$ with $Z_4$ symmetry have a similarly simple
solution, which however reproduces both the plasma frequency and
the Debye mass in the isotropic limit through the dispersion law
\be\label{N8Z4eql}
\omega^2={m^2(1+\xi)\03(1+\xi/3)^2}+{k^2\03}.
\ee
At larger momenta the solution becomes spacelike, whereas
the continuum dispersion law would approach the light-cone
exponentially in the timelike domain.

With increasing $\mathcal N$, the continuum result for
the longitudinal dispersion relations is
approximated from below (i.e.\ spacelike
momenta) by the mode with largest frequency. The Landau cut
at spacelike momenta is approximated by an increasing number
of poles, one of which is connected to the time-like
plasmon mode.\footnote{In the continuum, the longitudinal
plasmons have effectively a finite range of possible
momenta because the residue vanishes exponentially as the
light-cone is approached.}
For $\mathcal N=12$ and 20, there
are again two spacelike branches, now given by biquadratic
equations. The cases $\mathcal N=20$, $Z_5$ and $\mathcal N=12$, $Z_3$
again share the same equation, which reads
\be
45\omega^4-30\omega^2k^2+k^4=
{135m^2(1+\xi)\045+30\xi+\xi^2}
[5\omega^2(45+6\xi+\xi^2)-k^2(15+10\xi+3\xi^2)].
\ee

\begin{figure}
\includegraphics[width=2.5in]{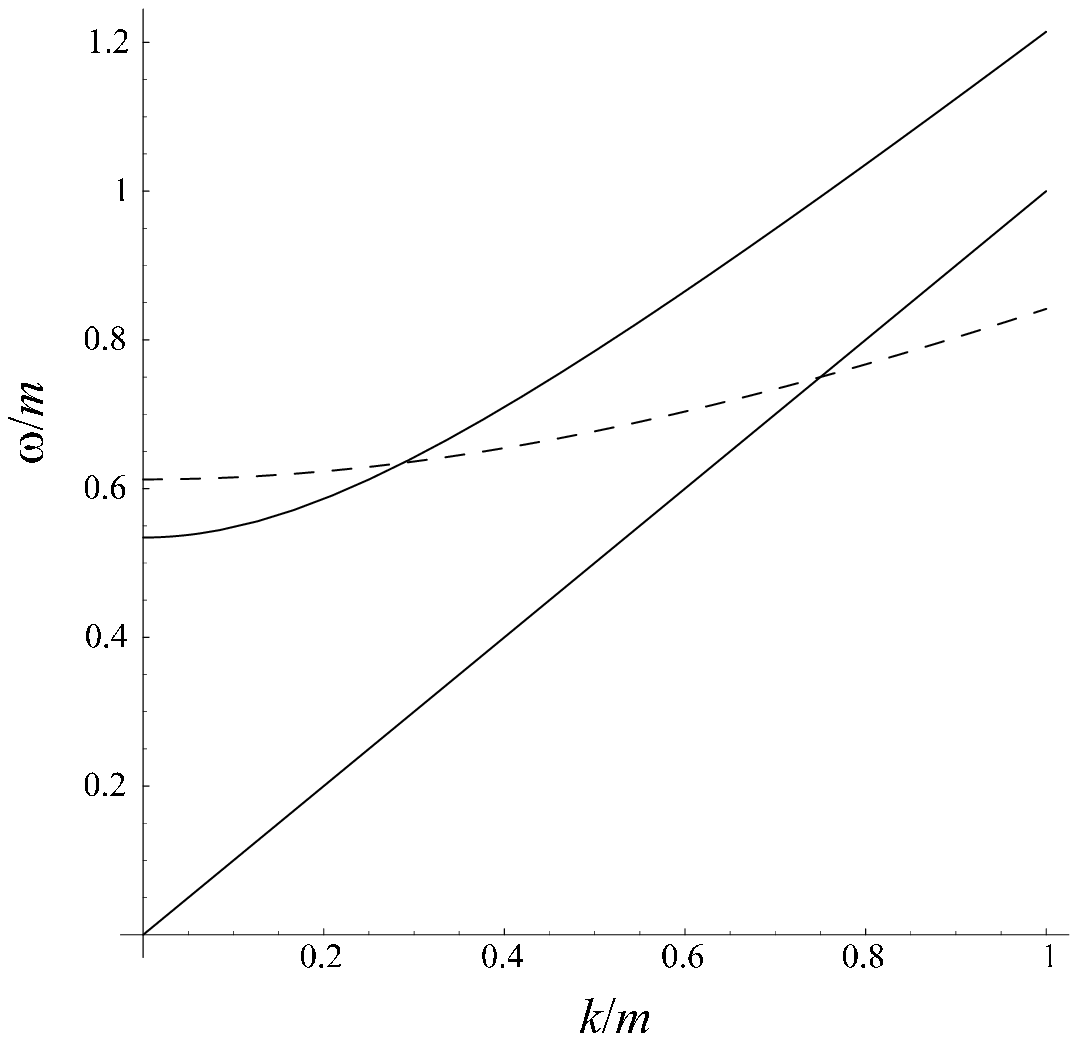}
\includegraphics[width=2.5in]{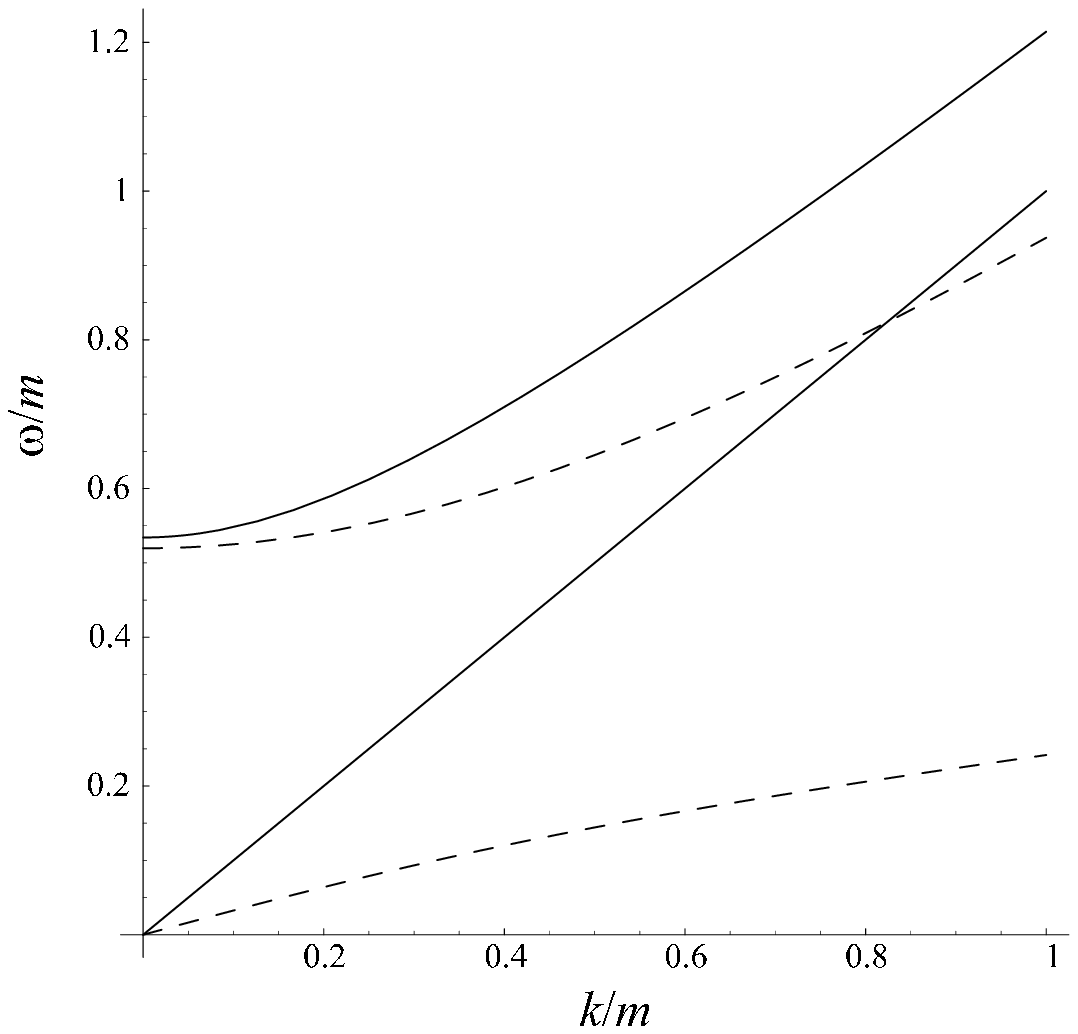}
\centerline{\hfil (a) $\mathcal N=8$ \hfil\hfil (b) $\mathcal N=20$ \hfil}
\caption{The longitudinal dispersion laws $\omega(k)$ for $\xi=1$
in the continuum (full lines) and for the
octahedron ($\mathcal N=8$, $Z_4$) and 
the icosahedron ($\mathcal N=20$, $Z_5$) (dashed lines).
\label{figdispl}}
\end{figure}

\section{1+1-dimensional lattice simulations}\label{sect:1D}

In the previous section we have seen that for oblate 
momentum distributions of plasma constituents, $\xi>0$, there
are unstable modes for $|k_z|<\mu$ and sufficiently small $k_\perp$.
In the linear regime, where field amplitudes $A$ are much smaller
than $k/g$, time evolution is determined by the above dispersion laws
and all modes evolve independently. Unstable modes grow exponentially
with growth rate $\gamma(\mathbf k)$, which for a given value of
$k_z$ is largest when $k_\perp=0$. We denote the scale of maximal growth
by $k_*$ and the corresponding maximal growth rate by $\gamma_*$.
This scale is of the same order of magnitude as the Debye screening
mass or the thermal mass of propagating plasmons, namely $\sim gp_{hard}$,
which at weak coupling is much larger than the color relaxation
scale $\sim g^2p_{hard}$ or the scale of large-angle scattering rates
$\sim g^4p_{hard}$.
This is what makes plasma instabilities a prime candidate for the
mechanism of fast isotropization in a weakly coupled plasma.

In the following we shall study the evolution of collective
soft fields starting from tiny random fluctuations. If the
initial field amplitudes are sufficiently small to give a large
number of e-folds in the linear regime, this will lead to
field configurations that are dominated by the modes of largest growth.
For $\xi>0$ this will favor modes with $|k_z|\approx k_*$ and
$k_\perp \approx 0$. The special
initial conditions with strictly $k_\perp=0$,
i.e.\ constant modes with respect to the spatial directions
transverse to the anisotropy direction
should therefore be an idealization of particular interest.
All fields are then functions of only one spatial coordinate,
and the equations of motion are simplified according to Eqs.~(\ref{1+1eoms}).

In Ref.~\cite{Arnold:2004ih}, non-Abelian dynamics
with such initial conditions was studied numerically
with a drastically simplified model for the induced current (\ref{Jind0}),
namely
\be\label{jAL}
j_\alpha^{\rm AL}=\mu^2 A_\alpha,\quad \alpha=x,y.
\ee
As was shown in Ref.~\cite{Arnold:2004ih}, this correctly reflects
the static limit of the anisotropic hard-loop effective
action for fields that vary only in the anisotropy direction, 
but it neglects its general frequency dependence and dynamical
nonlinearity. Already at the linearized level, the former
complication means that modes with vanishing wave vector, $\mathbf k=0$,
are stable, see Figs.~\ref{figgammas} and \ref{figgamma},
whereas the toy model (\ref{jAL}) implies a growth rate
$\gamma=\sqrt{\mu^2-\mathbf k^2}$, which is maximal at $\mathbf k=0$,
where it equals $\mu$. As Figs.~\ref{figgammas} and \ref{figgamma}
show, the anisotropic distribution function (\ref{faniso})
with $\xi>0$ leads to $\gamma_*<k_*<\mu$ instead.

In simulations of 
the toy model (\ref{jAL}), the authors of Ref.~\cite{Arnold:2004ih}
have found that unstable non-Abelian modes, which are
constant modes with respect to transverse space, might
behave very similarly to Abelian Weibel instabilities also
in the nonlinear regime. In the latter, they observed
rapid Abelianization, both locally and globally.

Using the $\mathbf v$-discretized equations of motion,
(\ref{DHLW})
and (\ref{DHLF}), we have extended the lattice simulations
of Ref.~\cite{Arnold:2004ih}
to the full hard-loop effective theory,
with the main findings presented  in Ref.~\cite{Rebhan:2004ur}.
In the 1+1-dimensional situation we found some modification
of the evolution of the instabilities when they first reach nonlinear
size. Subsequently, growth with rates close to the Abelian
growth rate was restored, however.

In the most recent paper \cite{Arnold:2005vb} a similar analysis
was performed, using a different discretization method for
the hard-loop effective theory. While confirming our
1+1-dimensional findings, they 
also considered random initial fluctuations
in all spatial directions and the resulting 3+1-dimensional evolution
with the conclusion that the late-time behavior, deeper in the
nonlinear regime, is in fact modified importantly. In this section
we shall concentrate on the 1+1-dimensional situation, giving
a detailed account of the results presented in Ref.~\cite{Rebhan:2004ur}.
After extending this also to gauge group SU(3), we shall
give our first results for a 3+1-dimensional simulation of
the $\mathbf v$-discretized hard-loop effective equations
in Sect.~\ref{sect:3D}.

Like Ref.~\cite{Arnold:2004ih} we work in temporal
axial gauge, $A^0=0$, and take initial conditions corresponding to
small random chromoelectric fields $\mathbf E=-\6_t \mathbf A$
with polarization transverse to the $z$-axis,
and all other fields vanishing, which
in our case includes the auxiliary fields
$\mathcal W_{\mathbf v}$. 
This initial condition satisfies the Gauss law constraint,
$
\mathbf D \cdot \mathbf E = {\mathcal N}^{-1}\sum_{\mathbf v} 
\mathcal W_{\mathbf v},
$
whose continued fulfilment is monitored in the simulation, but not
enforced. As a further nontrivial check of our numerics,
we tracked conservation of
the total energy (\ref{Etot}), 
stopping the simulation when this signalled loss of accuracy.
The lattice versions of the 1+1 dimensional equations of motion
that we have employed
are given in detail in Appendix \ref{applatt}.

In the 1+1-dimensional simulations below we used anisotropy
parameter $\xi=10$. We shall express all dimensionful quantities
by the  asymptotic mass of transverse excitations, whose
relation to the mass scale $m$ appearing in the
coefficients (\ref{avbvxi}) is given in Table~\ref{tablem}.
While the coupling constant $g$ could be absorbed by
redefinitions of our dynamical quantities, we shall make
their appearance explicit in our final results.

\begin{table}
\caption{Debye mass $m_{el}$, magnetic instability scale $\mu$,
wave vector $k_*$ and growth rate $\gamma_*$ for the
modes of maximal growth, transverse and longitudinal
plasma frequency, and the asymptotic mass
of transverse gluons for the isotropic case and
the anisotropic case assumed in the numerical simulations.
%, all in
%units of $m$ defined in Eq.~(\ref{m2norm})
\label{tablem}}
\begin{ruledtabular}
\begin{tabular}{c||r|r|r|r|r|r|r}
$\xi$ & $m_{el}/m$ & $\mu/m$ & $k_*/m$ & $\gamma_*/m$ & $\omega_{pl,t}/m$ &  $\omega_{pl,\ell}/m$ & $m_\infty/m$ \\
\hline\hline
0     & 1     & --    & --    & --         & 0.577350 & 0.577350 & 0.707107 \\
\hline
10    &1.64296&1.07225&0.398025&0.120289& 0.339075 &0.412228 &0.447144 \\
\end{tabular}
\end{ruledtabular}
\end{table}

The lattice spacing used in the simulations
presented in this section was chosen such that
$a=0.0707 m_\infty^{-1}$. We have taken a one-dimensional spatial
lattice with periodic boundary conditions and 
5,000 sites, so that the physical size
is $L\approx 350 m_\infty^{-1}$. We shall show results
for the icosahedron $\mathcal N=20, Z_5$, and
disco-ball discretizations $\mathcal N=100$, 800, and 2000,
with $N_z\times N_\varphi=20\times 5$, $50\times 16$, and $100\times 20$, 
respectively. In the leapfrog algorithm specified in Appendix
\ref{applatt} we used time steps $\epsilon/a$ from $1/100$ to
$1/250$. The initial conditions are given in Eq.~(\ref{initcond1d}),
and the choice of the parameter $\sigma$ therein corresponds
to starting with a random initial seed chromoelectric field of
root-mean-square amplitude $0.012\,m_\infty^2/g$.

In the numerical simulations, the Gauss law constraint, which
is satisfied by the initial conditions, remained satisfied within
machine accuracy throughout. As a further check, we monitored the
total energy (\ref{Etot}) and stopped the simulation when
this had O(1) violations. For all but the largest times, this
quantity was conserved within less than 1\%; only at the very end
when the energy in the various field components had grown by
factors larger than $10^6$ was there energy violation at the
percentage level, which quickly thereafter exploded, signalling
uncontrollable discretization errors.

\subsection{Currents and energies}

\begin{figure}[thb]
\vspace{1cm}
\includegraphics[width=6.5in]
{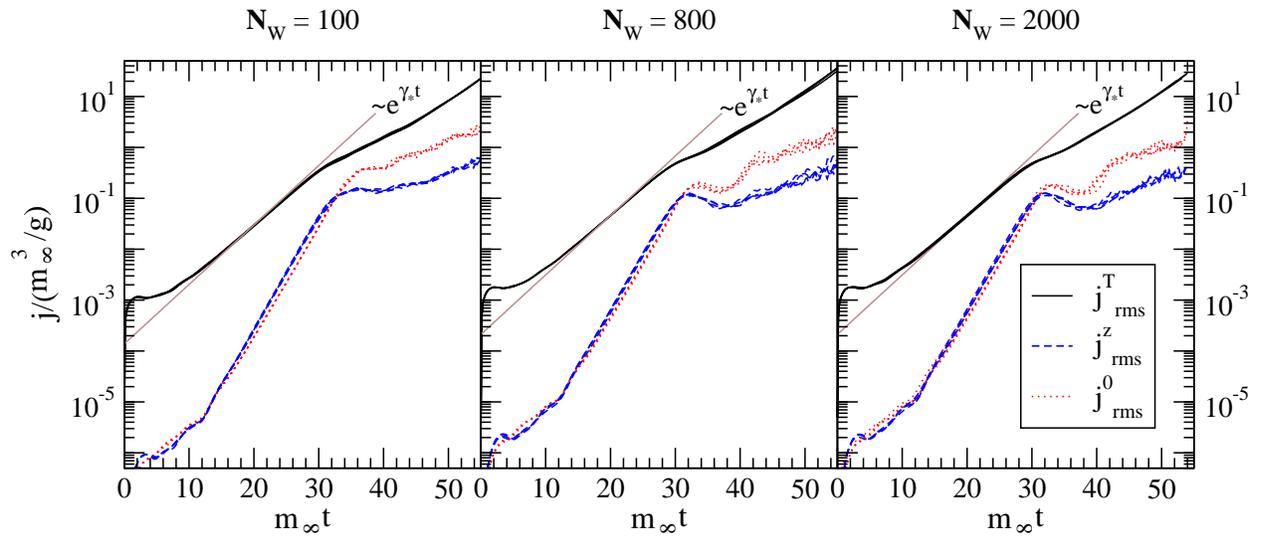} 
\caption{The average current $j_{rms}^\perp$ (black),
$j_{rms}^z$ (blue) and rms charge density $j^0_{rms}$ (red)
for $N_W\equiv\mathcal N=100,800,2000$ using a collection of 4 runs each with
different random initial conditions.
\label{figjrms}}
\end{figure}

Fig.~\ref{figjrms} shows the evolution of the
average root-mean-square transverse and longitudinal
currents and charge density defined by
\be
j_{rms}^{\perp,z,0}=
\left[ \int_0^L {dz\0L} 2 \tr ((\mathbf j^{\perp,z,0})^2) \right]^{1/2},
\ee
comparing the cases $\mathcal N=100$, 800, and 2000.
The transverse current 
grows exponentially with a growth rate that is most of the
time only slightly below $\gamma_*$, %the Abelian value, 
except for a transitory reduction at the beginning
of the nonlinear regime, when $j_{rms}\sim m_\infty^3/g$. 
Our initial condition (only nonvanishing transverse electric fields at
$t=0$) give rise to longitudinal currents and charge densities
only through non-Abelian interactions (in the Abelian regime all
modes would decouple and evolve independently according to
the linearized dispersion laws).
The non-Abelian interactions naturally lead to growth rates
for $j^z$ and $j^0$ which are double that of $j^\perp$, and
such a behavior is seen in Fig.~\ref{figjrms} after some
initial delay caused by the fact that the unstable modes first have
to grow out of the initial random
mixture of stable and unstable ones.
The different cases $\mathcal N=100$, 800, and 2000 lead to
only slightly different behavior in the nonlinear regime, where
higher $\mathcal N$ seems to be required to capture the
precise variations of the subdominant components $j^z$ and $j^0$. 

\begin{figure}
\vspace{1cm}
\includegraphics[width=6.3in]%,angle=-90]
{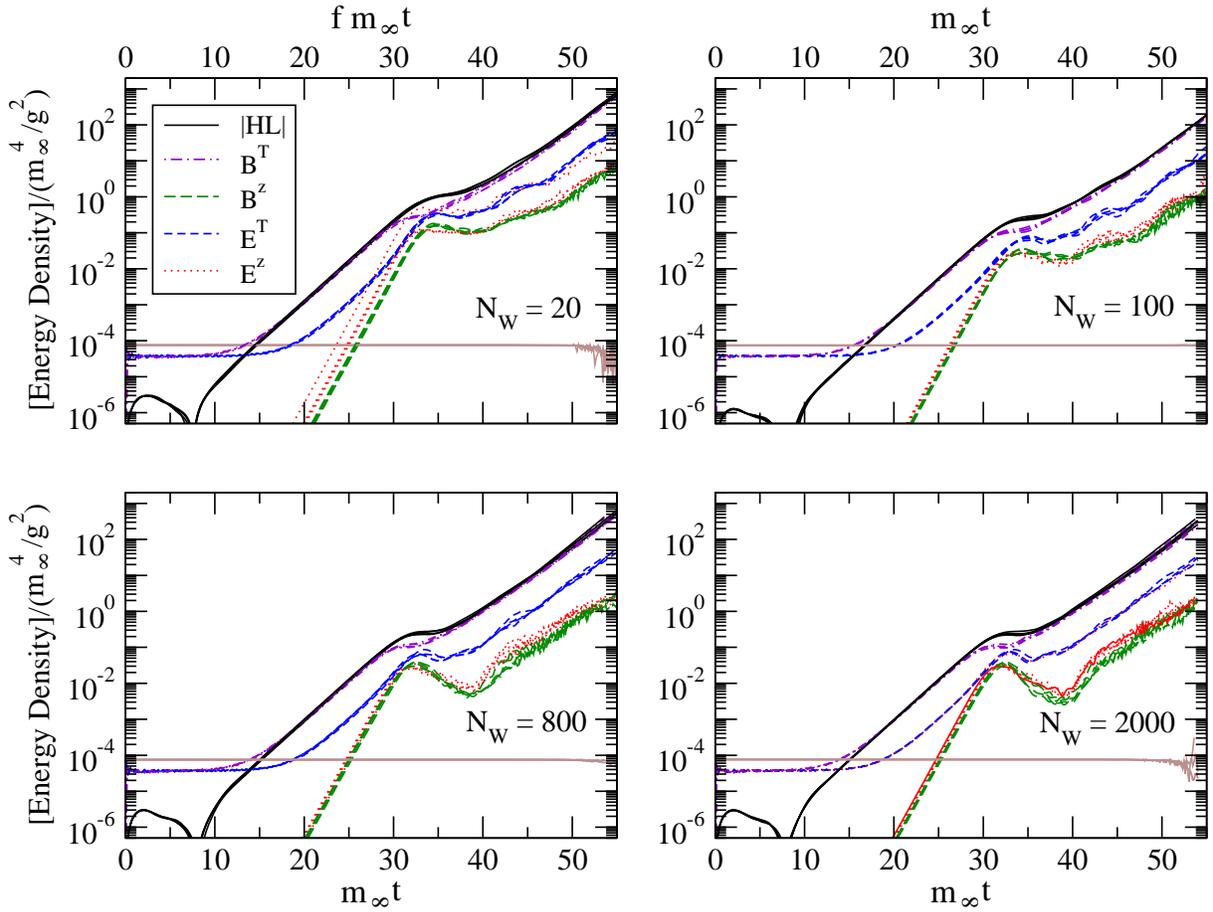}
\caption{Average energy densities $\mathcal E$
in transverse/longitudinal chromomagnetic/electric
fields and the total energy density contributed by
hard particles, $\mathcal E({\rm HL})$,
for various values of $\mathcal N$. 
In the case $\mathcal N=20$,
where there is a noticeable difference between $\gamma_*$ at
the discretized level from the continuum value (see middle plot
in Fig.~\ref{figgammas}), we have rescaled the factor $m_\infty$
in front of $t$ by $f=\gamma_*(\mathcal N=20)/\gamma_*=1.29336$.
The horizontal line terminating in
fluctuations is the total conserved energy (\ref{Etot}).
\label{figNen}}
\end{figure}

Closely related to this,
Fig.~\ref{figNen} shows how the exponentially growing energy 
that is transferred
from hard to soft scales gets distributed among chromomagnetic
and chromoelectric collective fields. The dominant contribution
is in transverse magnetic fields, and it grows roughly with
the maximum rate $\gamma_*$ both in the linear as well as in
the highly nonlinear regime, with a transitory slowdown in between. 
Transverse electric fields behave similarly,
and are suppressed by a factor of the order of $(\gamma_*/k_*)^2$.
Notice that in the
model of Ref.~\cite{Arnold:2004ih} the situation
would be different: Because $k_*$ is zero there, the dominant
energy component is then from transverse electric fields, whereas the
relative importance of magnetic fields drops with time.
In the 1+1-dimensional evolution, the transverse magnetic field
component is the dominant one, both in the linear and in the nonlinear
regime.

The appearance of
longitudinal contributions, which are absent in the initial conditions
we have chosen, is again a purely non-Abelian effect. While 
completely negligible at first,
they have a growth rate which is double the one in the transverse sector,
and they begin to catch up with the latter when
the nonlinear regime is reached first. 
At this point, the general growth stalls for a time of order $\gamma_*^{-1}$.
When the transverse magnetic field resumes its growth (with
the transverse electric field following with some delay), the 
respective longitudinal components drop for some time before
also starting to grow again. This dip in the subdominant 
longitudinal components
turns out to depend on $\mathcal N$, whereas the overall behavior
is already reproduced well by the rather small value of $\mathcal N=20$
(the icosahedron).

Deeper into the nonlinear regime, the (transversely constant) unstable
modes grow at roughly equal rates. There the results are only
weakly dependent on the degree of hard-loop discretization, $\mathcal N$.

\subsection{Color correlation}

Had we assumed initial conditions where all fields would point
in the same color direction all over the spatial lattice,
we would have found only strictly exponential growth once
the stable modes have become of negligible importance.
The behavior would then be exactly the same as with
Abelian Weibel instabilities, which grow exponentially until
they come into conflict with the assumptions of the hard-loop approximation
(namely that the fields have only small effects on the trajectories
of the hard particles), whereupon isotropization is supposed to set in.

Apart from a brief transitory slowdown, the 1+1-dimensional simulations
thus evolve similarly to Abelian instabilities 
both in the linear and the strongly nonlinear
domain (at least as far as the dominant transverse
components are concerned). 
A measure of local ``non-Abelianness'' can be defined
by \footnote{%
This definition coincides with the definition in
Ref.~\cite{Arnold:2004ih} in the simplifying case (\ref{jAL}), but
is gauge invariant in the full theory
also when the restriction to 1+1 dimensional
configurations is removed.}
\be
\bar C[j] = \int_0^L {dz\0L} { \left\{ \tr \left( (i[j_x,j_y])^2 \right) 
\right\}^{1/2}\0 \tr (j_x^2+j_y^2) }.
\ee
Indeed, when the nonlinear regime is entered, this quantity is
found to decay roughly exponentially (with some hiccup brought out
only with sufficiently large $\mathcal N$), but not as fast as
it was found in the toy model of Ref.~\cite{Arnold:2004ih}.

\begin{figure}
\vspace{1cm}
\includegraphics[width=6.3in]
{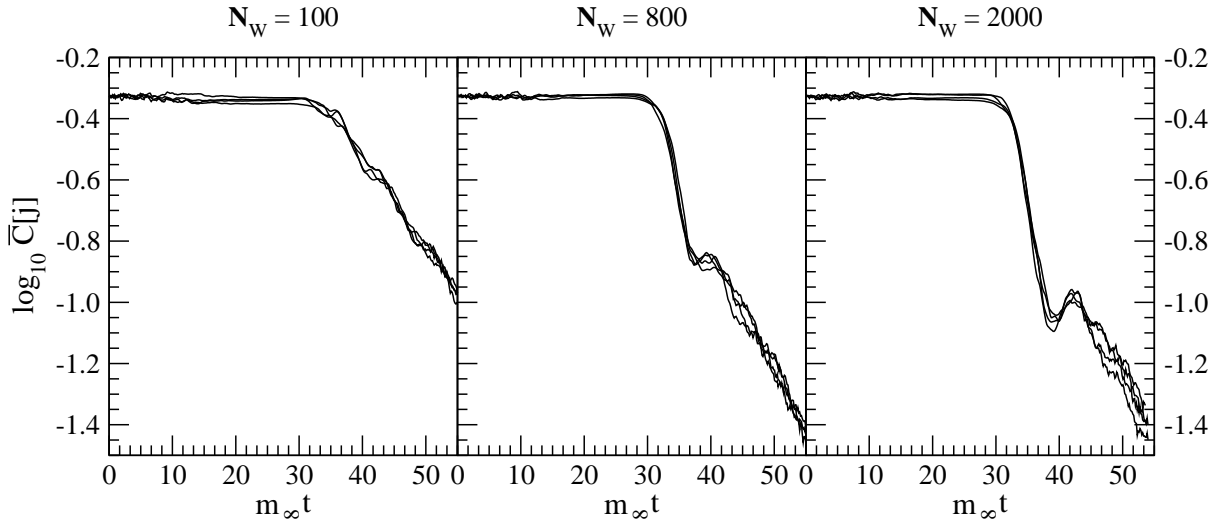}   
\caption{The 
relative rms size $\bar C$ of commutators $[j_x,j_y]$ as a function of time
for $\mathcal N=100,800,2000$.
\label{figCbars}}
\end{figure}

Ref.~\cite{Arnold:2004ih} also observed a concurrent global
Abelianization, by comparing the correlation among parallel
transported spatially separated commutators to the correlation
of parallel transported fields. 
Generalizing the quantity used in Ref.~\cite{Arnold:2004ih},\footnote{%
We use currents instead of
transverse gauge field components,
because they transform like adjoint matter when the
restriction to 1+1-dimensional situations is removed.}
we define
\be
\chi_A(\xi)={N_c^2-1\02N_c}
\int_0^L {dz\0L} { \tr \left\{
(i[j_i(z+\xi),\mathcal U(z+\xi,z)j_j(z)])^2 \right\}
\0 \tr\{ j_k^2(z+\xi) \} \tr \{ j_l^2(z) \} }
\ee
where $\mathcal U(z',z)$ is the adjoint-representation parallel transport
from $z$ to $z'$. When colors are completely uncorrelated over a
distance $\xi$, this quantity equals unity; if they point in the
same direction, this quantity vanishes. Following Ref.~\cite{Arnold:2004ih} 
we define the ``Abelianization correlation length''
$\xi_A$ as the smallest distance where $\chi_A$ is larger than 1/2,
\be
\xi_A[j]=\min_{\chi_A(\xi)\ge 1/2}  (\xi).
\ee

This we compare with a general correlation length, not focussing on
color, defined through the gauge invariant function 
\be
\chi(\xi)={ \int_0^L {dz\0L} \tr\{ j_i(z+\xi)
\mathcal U(z+\xi,z)j_i(z)\} \0
\int_0^L {dz\0L} \tr\{ j_l(z) j_l(z)\} }.
\ee
This function now vanishes when fields are uncorrelated over a distance
$\xi$, and it is normalized such that $\chi(0)=1$.
We thus define the general correlation length through
\be
\xi[j]=\min_{\chi(\xi)\le 1/2}  (\xi) .
\ee

\begin{figure}
\includegraphics[width=6.3in]
{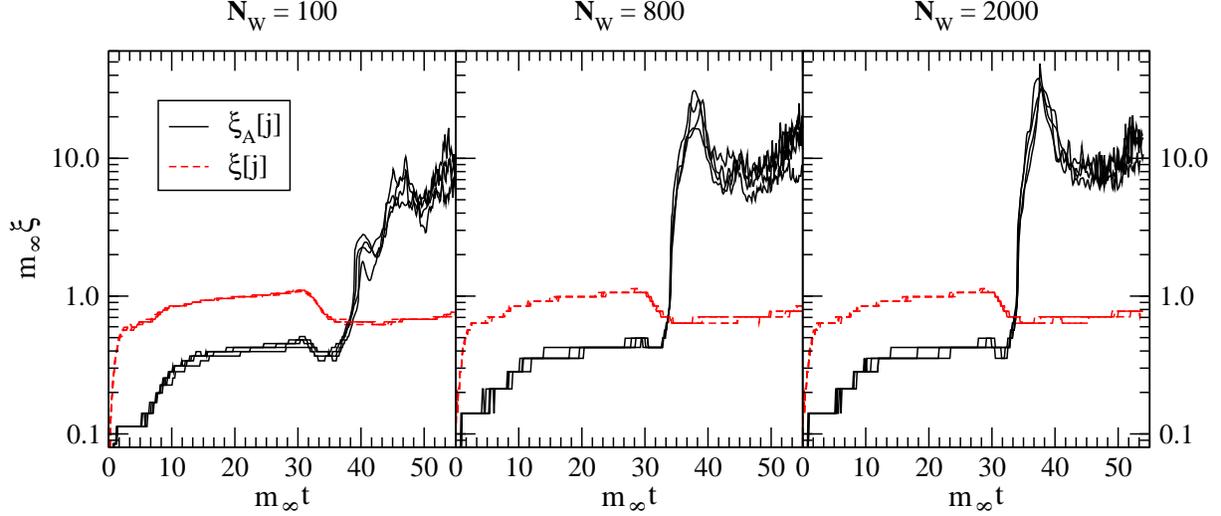}  
\caption{The Abelianization correlation length $\xi_A[j]$ compared to the
ordinary
correlation length $\xi[j]$ for $\mathcal N=100,800,2000$.
\label{figxias}}
\end{figure}

The evolution of the general and the Abelianization correlation length
is shown in Fig.~\ref{figxias}. In contrast to 
the toy model of Ref.~\cite{Arnold:2004ih}, where $\xi_A$ was found to
rapidly grow to full lattice size when $\bar C$ begins its decay
(while no such behavior was seen in general correlation lengths),
we find that Abelianization takes place only over domains of
limited sizes, see Fig.~\ref{figxias}. Deeper in the nonlinear
regime, $\xi_A$ is seen to have magnitudes comparable to the
wavelength of the modes of maximal growth,\footnote{Note again that in the
model of Ref.~\cite{Arnold:2004ih} $k_*$ vanishes,
which is presumably responsible for the different global behavior.}
$\lambda_*=2\pi/k_* \approx 7.06 \, %7.05866
m_\infty^{-1}$. 

In this regime, the different simulations with
different $\mathcal N$ agree fairly well, only in the
transition region between linear and nonlinear regime it is
essential to have higher values of $\mathcal N$ to bring out
an initial peak in the evolution of $\xi_A$.

\subsection{Extension to gauge group SU(3)}

\begin{figure}
\vspace{7mm}
\includegraphics[width=4in]%,angle=-90]
{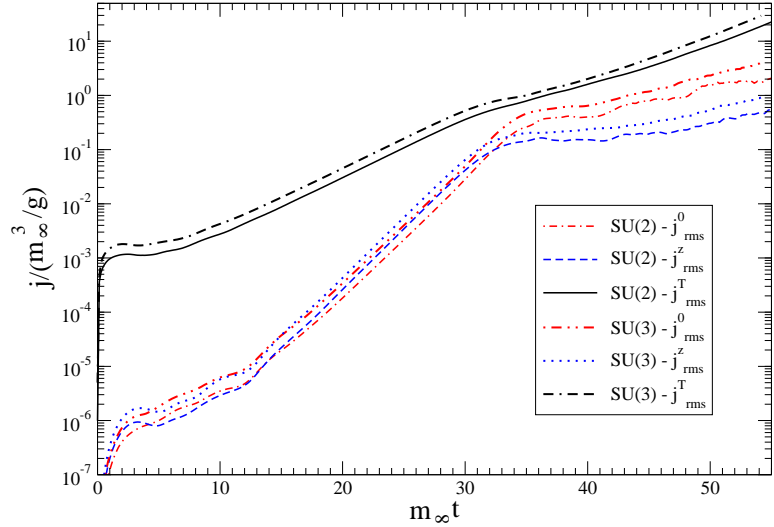}  
\caption{The average current $j_{rms}^\perp$,
$j_{rms}^z$ and rms charge density $j^0_{rms}$
with gauge group SU(3) and SU(2),
for one $\mathcal N=100$ run each. 
\label{figjrms3}}
\end{figure}

\begin{figure}
\vspace{10mm}
\includegraphics[width=5.5in]%,angle=-90]
{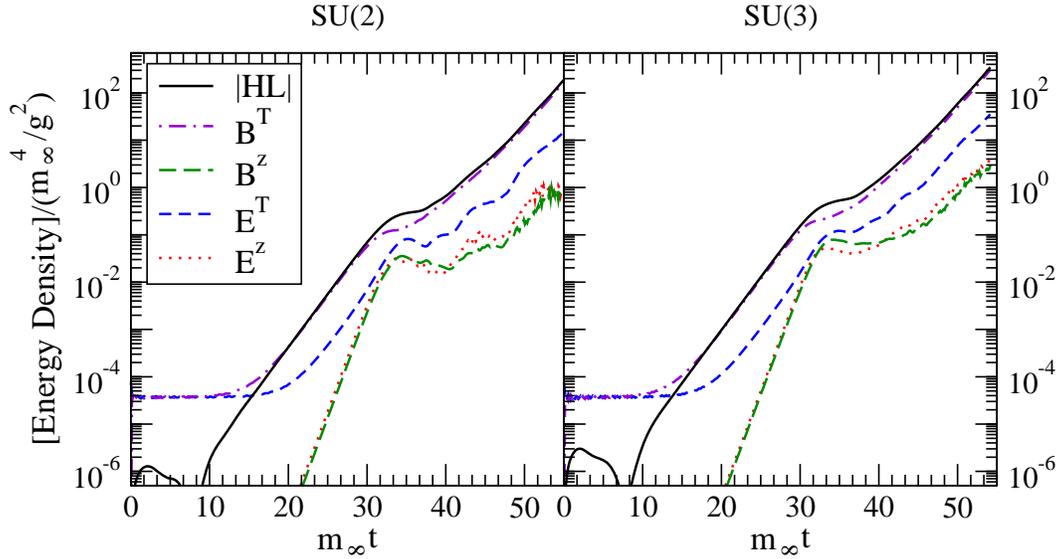}
\caption{SU(3) vs.\ SU(2) comparison of the
average energy densities $\mathcal E$
in transverse/longitudinal chromomagnetic/electric
fields and the total energy density contributed by
hard particles, $\mathcal E({\rm HL})$.
\label{figNen2v3}}
\end{figure}

\begin{figure}
\vspace{7mm}
\includegraphics[width=4in]%,angle=-90]
{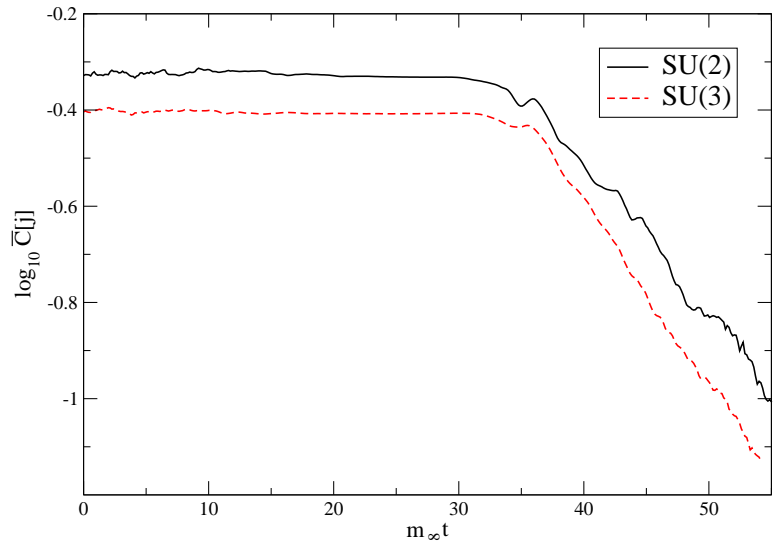} 
\caption{SU(3) vs.\ SU(2) comparison of the 
local Abelianization measure 
$\bar C$ of commutators $[j_x,j_y]$. 
\label{figCbars3}}
\end{figure}

\begin{figure}
\vspace{7mm}
\includegraphics[width=4in]%,angle=-90]
{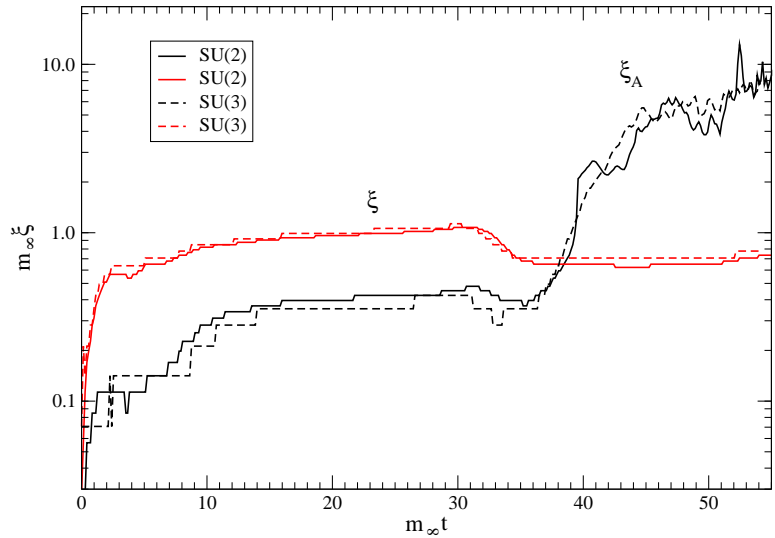}  
\caption{SU(3) vs.\ SU(2) comparison of
the Abelianization correlation length $\xi_A[j]$ 
together with
the ordinary
correlation length $\xi[j]$.
\label{figxias3}}
\end{figure}

The above numerical simulations have been performed with
gauge group SU(2), which is the most economical in computer time.
There is no particular reason why non-Abelian plasma
instabilities should strongly depend on the gauge group
(in particular when local Abelianization occurs).
Nevertheless, this needs to be checked.
In this section we discuss the changes which occur when the gauge group
is taken to be SU(3) instead of SU(2).  Functionally, the algorithm is
exactly the same with the only difference being the underlying matrix
representation for the fields.  Figures \ref{figjrms3} -- \ref{figxias3} 
show the effect of upgrading the gauge group SU(2) to SU(3) 
for $\mathcal N=100$.

In Figs.~\ref{figjrms3} and \ref{figNen2v3} we compare
the results for the various current and energy components
in the case of gauge group
SU(3) with those for SU(2), both with $\mathcal N=100$.
Only minor differences are seen, with a little higher
energies being reached in the nonlinear regime in SU(3),
but that would need more extensive tests to be taken as
a conclusion.

The amount of local and global Abelianization is shown
in Figs.~\ref{figCbars3} and \ref{figxias3}, respectively.
Again, only small differences can be observed.

\section{3+1-dimensional lattice simulations}\label{sect:3D}

We finally turn to the question to what extent 
relaxing strict translation invariance
in transverse directions, which led to the great simplification
of effectively 1+1-dimensional dynamics, does modify the
evolution of plasma instabilities. In the linear regime,
the time evolution favors the modes which are constant
with respect to transverse directions, because these are
the ones of largest growth. Thus, if one starts out with
arbitrarily small seed fields, transverse translational
symmetry will be arbitrarily well prepared for the subsequent
nonlinear evolution (after correspondingly arbitrarily
long Abelian evolution of course). This is the case that was covered by
the previous section. However, generic initial conditions
that take a finite time to evolve into the nonlinear
regime (or start out there) 
will always have some breaking of transverse
translational symmetry, and then
one has to simulate full 3+1 dimensional dynamics to take
this into account.

\begin{figure}[h!tb]
\vspace{7mm}
\includegraphics[width=4in]%,angle=-90]
{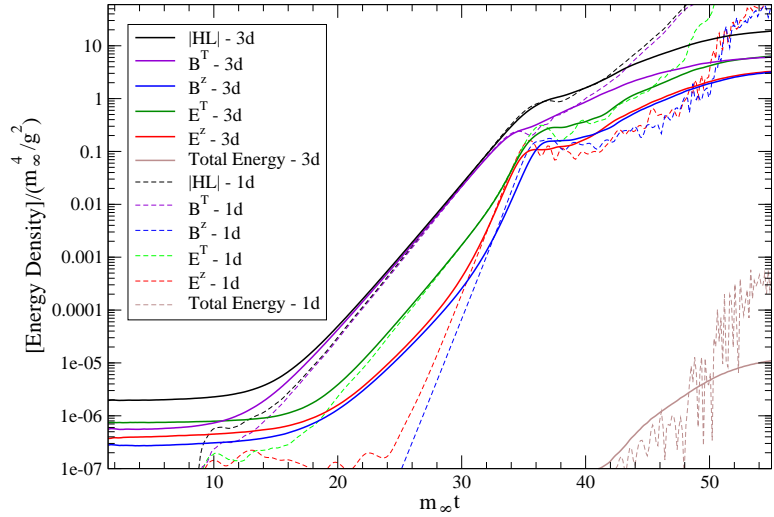}
\caption{Average energy densities $\mathcal E$
in transverse/longitudinal chromomagnetic/electric
fields and the total energy density contributed by
hard particles, $\mathcal E({\rm HL})$
for 3+1-dimensional $\mathcal N=20$.
\label{figNenlog31}}
\end{figure}

\begin{figure}[h!tb]
\vspace{7mm}
\includegraphics[width=4in]%,angle=-90]
{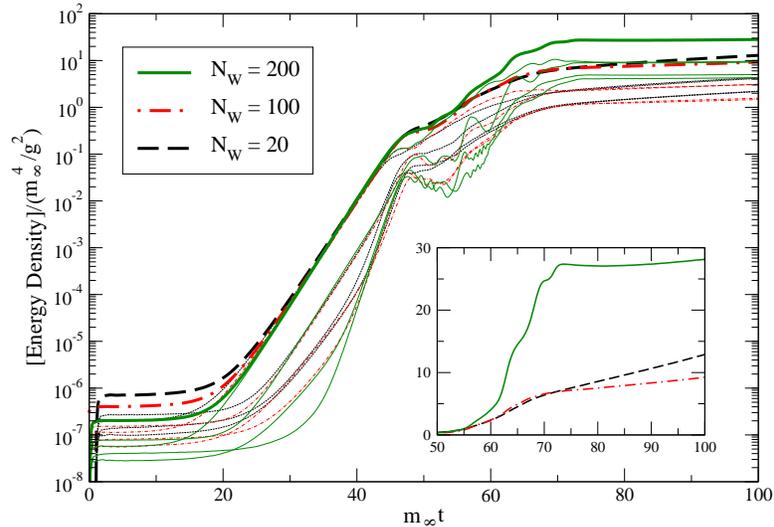}
\caption{Comparison of average energy densities $\mathcal E$
for 3+1-dimensional simulations with $\mathcal N=20,100,200$
on $96^3, 88^3, 69^3$ lattices.
Thick lines correspond to $\mathcal E({\rm HL})$ and 
thin lines are the components which can be inferred from
Fig.~\ref{figNenlin31}.
In the case $\mathcal N=20$ we have rescaled $m_\infty \rightarrow f\,m_\infty$
with $f=\gamma_*(\mathcal N=20)/\gamma_*=1.29336$ in order to properly normalize the
results for comparison.  In addition, we shifted the $\mathcal N=20$ data 0.9 units
in $m_\infty t$ to the right and the $\mathcal N=100$ by 0.1 units
in $m_\infty t$ to the left in order to 
roughly align the exponential-growth phase in the Abelian regime.
%obtain roughly the same abelian behavior.
Inset shows late-time behavior of the hard-loop energy density with a linear
scale.
\label{figNenlog32}}
\end{figure}

\begin{figure}[h!tb]
\vspace{7mm}
\includegraphics[width=4in]%,angle=-90]
{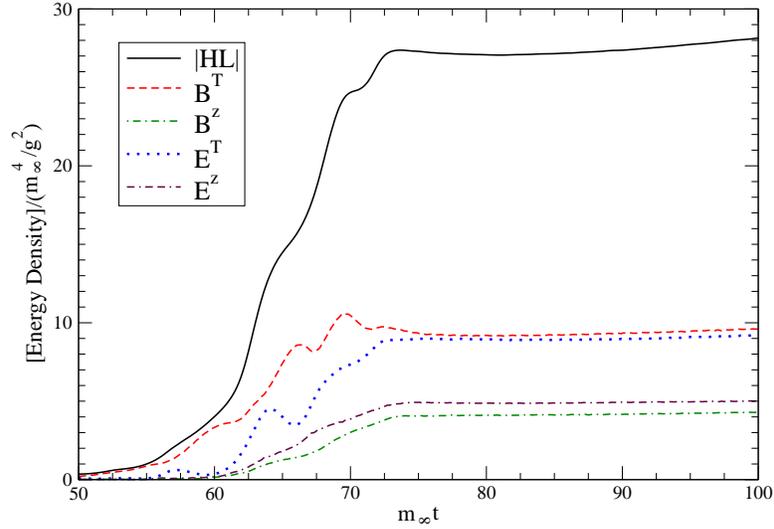}
\caption{Average energy densities $\mathcal E$
in transverse/longitudinal chromomagnetic/electric
fields and the total energy density contributed by
hard particles, $\mathcal E({\rm HL})$
for 3+1-dimensional $\mathcal N=200$, in linear scale.
\label{figNenlin31}}
\end{figure}

\begin{figure}[h!tb]
\vspace{7mm}
\includegraphics[width=4in]%,angle=-90]
{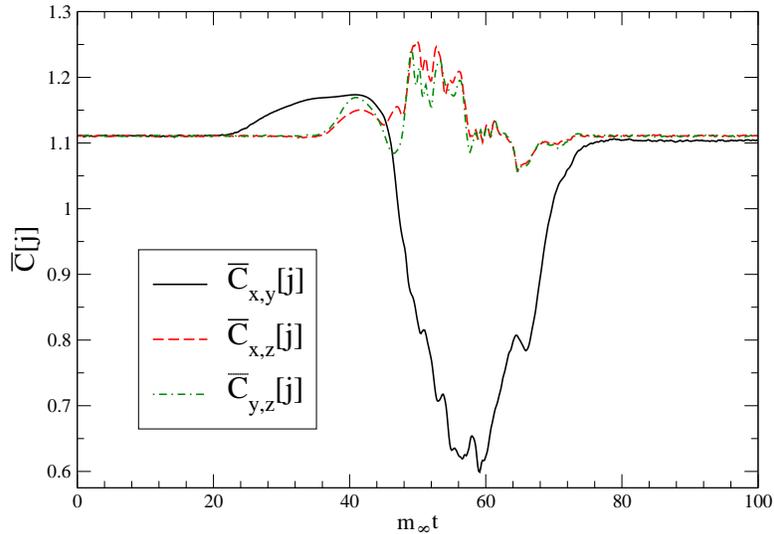}
\caption{The relative rms size of commutators ${\bar C}^2_{i,j}={\rm tr} ([j_i,j_j])^2/({\rm tr} j_i^2\,{\rm tr} j_j^2)$ as a function of time compared for 3+1-dimensional versus 1+1-dimensional evolution with $\mathcal N=200$.
\label{figcbarj31}}
\end{figure}

Figures~\ref{figNenlog31}--\ref{figcbarj31} show
our first results of such 3+1-dimensional lattice simulations.
In these simulations we have no longer set up initial
conditions corresponding only to transverse electric fields,
but formulated these initial conditions in terms of the 
$\mathcal W$ fields, see Eq.~(\ref{3dinitcond}).

We have investigated both asymmetrical and symmetrical lattices,
but here we only present results from symmetric ones.
In Fig.~\ref{figNenlog31} we plot the resulting energy densities 
for a $96^3$ lattice and the icosahedral approximation, $\mathcal N=20$,
compared to a one-dimensional simulation using the same parameters. As
can be seen from this figure, differences from the one-dimensional
simulations begin to appear at $m_\infty t \sim 35$ which is when
the field amplitudes have reached order $m_\infty/g$. 
The chosen lattice spacing corresponds to $a=0.28m_\infty^{-1}$
so that the physical volume of the lattice is roughly $(28 / m_\infty)^3$.
In Fig.~\ref{figNenlog32} we compare the energy densities obtained 
using ${\cal N} = 20, 100, 200$, 
namely the icosahedron and disco-balls with $\mathcal N_z
\times \mathcal N_\varphi= 20 \times 5$ and $25\times 8$,
with lattice sizes $96^3$, $88^3$, and $69^3$,
respectively.  The thick lines are the energy
extracted from the hard particles (HL energy) and the other
various components of the total energy are shown for comparison
but not labelled explicitly.  As can be seen from this figure 
there is general agreement between all three discretizations, the 
chief difference being that for late times the then approximately 
linear growth rate 
seems to decrease with increasing ${\cal N}$.  The fluctuations
in the overall amplitude are due to the different random initial
conditions used for each run.

In the linear regime, i.e.\ field amplitudes $\ll m_\infty/g$,
the differences are only due to the different amount of
initial energy densities in longitudinal and transverse components.
In both the 3+1-dimensional and the 1+1-dimensional case,
the dominant component is from transverse magnetic fields,
followed by transverse electric fields (suppressed by a factor
of $(\gamma_*/k_*)^2<1$. Longitudinal components are initially
absent in the 1+1-dimensional preparation of the system, but
not in the full 3+1-dimensional case because of the
different initial conditions (\ref{3dinitcond}). Nevertheless, at the later
stages of the linear evolution, they behave rather similarly.
The regime where specifically non-Abelian effects come into the
play first is also very similar in the two cases, and remain
so for a while. However, whereas the 1+1-dimensional modes
eventually recover their initial growth rates of the linear regime,
the transversely nonconstant modes seem to be the determining
factor deeper into the nonlinear regime, and these behave
rather differently from the former. In Figs.~\ref{figNenlog31} and \ref{figNenlog32}
one sees subexponential behavior at late times,
which in fact seems to tend to a linear growth at large times.
This can be seen most clearly in Fig.~\ref{figNenlin31} where we
plot the components of the energy density at late times on a 
linear scale for a $69^3$ lattice and $\mathcal N=200$.
The chosen lattice spacing corresponds to $a=0.28m_\infty^{-1}$
so that the physical volume of the lattice is roughly $(19 / m_\infty)^3$.

This finding is in agreement with the most recent results
by Arnold, Moore, and Yaffe \cite{Arnold:2005vb}, using
a different discretization method for the hard-loop effective
theory (namely truncation of expansions in spherical harmonics).
While Ref.~\cite{Arnold:2005vb} presented a very extensive
analysis of possible systematic uncertainties, our results,
obtained with cruder discretizations, are still somewhat
preliminary and require further
systematic studies. Nevertheless, we already find agreement also in 
some of the finer
details, such as the relative importance of longitudinal
field components in the late time evolution.

In Fig.~\ref{figcbarj31} we moreover display results for the normalized
size of the commutators ${\bar C}^2_{i,j}={\rm tr} ([j_i,j_j])^2/({\rm tr} j_i^2\,{\rm tr} j_j^2)$
which give a measure of local Abelianization with respect to the three 
possible spatial directions.  As can be seen from this figure there is
initially Abelianization in the $xy$ commutator starting at $m_\infty t \sim 40$,
while the other two commutators show no evidence for Abelianization.  At later
times ($m_\infty t >\sim 60$),
the $xy$ commutator becomes less and less abelian and at late times, in the
linear-growth phase,
the soft-current commutators are nearly isotropic.

\section{Conclusion and Outlook}\label{sect:Concl}

In this paper we have presented detailed results from our numerical
studies of discretized hard loops applied to the time-evolution of quark-gluon
plasma instabilities due to momentum space anisotropy.  
We extended our previous effective 1+1 dimensional analysis to include the
SU(3) gauge group finding that the results obtained 
for SU(3) were qualitatively similar to those of our SU(2) simulations.  
In addition, we have presented first results for the full 3+1 
dimensional evolution of the system using our 
method to discretize the hard-loop effective theory.  
We tentatively confirm the results of Arnold, Moore, and Yaffe 
\cite{Arnold:2005vb} who find linear
instead of exponential growth of the hard-loop energy at late times. 

The fact that 3+1 dimensional nonabelian plasma
instabilities seem to nearly saturate in the nonlinear regime where field
amplitudes are of order $m_\infty/g$ makes it even more interesting
and also self-consistent to study their dynamics and their role
in isotropization and thermalization of quark-gluon plasma
within discretized hard-loop approximations.
A more extensive and thorough study of 3+1 dimensional
quark-gluon-plasma instabilities will be the subject of
a forthcoming publication.

\acknowledgments

We thank P.~Arnold, D.~B\"odeker, M.~Laine, G.~Moore, and L.~Yaffe 
for valuable discussions.
M.S. was supported by the Academy of Finland, contract no. 77744.
P.R. was supported by the DFG-Forschergruppe EN 164/4-4.

\appendix

\section{Small $\xi$ expansion of dispersion laws}

\subsection{Continuum results}

The small-$\xi$ expansion of the continuum
transverse plasma frequency (\ref{ompltconti}) reads
\be\label{ompltcont}
\omega^2_{pl,t}/m^2=
\underbrace{1\03}_{p_1}-\underbrace{2\015}_{p_2}\xi+\underbrace{3\035}_{p_3}\xi^2+O(\xi^3).
\ee
(The coefficients $p_k$ are compared with the $\mathbf v$-discretized
results in Table \ref{tableA}). 

Propagating modes with small $k$ are characterized by
\be\label{omkcont}
\omega^2_t(k)=\omega^2_{pl,t}+\Bigl[\underbrace{5\06}_{q_1}
+\underbrace{19\0175}_{q_2}\xi+O(\xi^2)\Bigr]k^2+O(k^4).
\ee

The small-$\xi$ expansion of the 
asymptotic mass (\ref{masconti}) of the transverse excitations reads
\be\label{mascont}
m^2_\infty/m^2=\underbrace{1\02}_{r_1}-\underbrace{1\06}_{r_2}\xi
+\underbrace{1\010}_{r_3}\xi^2+O(\xi^3).
\ee

The plasma frequency in the longitudinal sector, Eq.~(\ref{ompllconti}),
has the expansion
\be\label{ompllcont}
\omega^2_{pl,\ell}/m^2=
\underbrace{1\03}_{s_1}-\underbrace{1\015}_{s_2}\xi
+\underbrace{1\035}_{s_3}\xi^2+O(\xi^3).
\ee

\subsection{Discretized hard loops}
\label{appdhlcomp}

\begin{table}
\caption{A comparison of parameters characterizing the
dispersion laws in anisotropic plasmas at small anisotropy
parameter $\xi$ between continuum and $\mathbf v$-discretized
hard loops. The upper part of the table refers to discretization by
regular polyhedra $\mathcal N=6,8,12,20$
with different orientations and hence
different discrete ($Z_\varphi$) rotational symmetry around the $z$-axis and
different numbers $N_z$ of distinct values $v_z$. 
The numbers on the right-hand side give the factors by which the coefficients
defined in (\ref{ompltcont}), (\ref{omkcont}), (\ref{mascont}), (\ref{ompllcont})
differ from the exact ones. 
These factors are given only up to the first term where there is a
deviation from the exact result.
The lower entries correspond to the ``disco-ball'' discretization
(\ref{disco}) with
regular spacing in $z$ and $\varphi$ (there the results actually
only depend on $N_z$ if $N_\varphi\ge2$).
\label{tableA} }
\begin{ruledtabular}
\begin{tabular}{c|cc||ccc|cc|ccc||ccc}
%\hline
$\mathcal N$ & $Z_\varphi$ & $N_z$ & $p_1$ & $p_2$ & $p_3$ & $q_1$ & $q_2$ & $r_1$ & $r_2$ & $r_3$ & $s_1$ & $s_2$ & $s_3$ \\
\hline\hline
6 & 4 & 3 & 1 & 0 &  & 0 &  & 1.50 & 0 & & 1 & 5.00 \\
6 & 3 & 2 & 1 & 1.67 & & 1.11 & & 1 & 1 &  0.56 & 1 & $-1.67$ \\
\hline
8 & 4 & 2 & 1 & 1.67 & & 1.11 & & 1 & 1 &  0.56 & 1 & $-1.67$ \\
8 & 3 & 4 & 1 & 0.56 & & 0.93 & & 0.75 & &  & 1 & 2.78 \\
\hline
12 & 5 & 4 & 1 & 1 & 0.47 & 1 & 1.84 & 0.83 & &  & 1 & 1 & 4.20 \\
12 & 3 & 4 & 1 & 1 & 1.30 & 1 & 0.53 & 1 & 1 & 1  & 1 & 1 & $-0.78$ \\
\hline 
20 & 5 & 4 & 1 & 1 & 1.30 & 1 & 0.53 & 1 & 1 & 1  & 1 & 1 & $-0.78$ \\
20 & 3 & 6 & 1 & 1 & 0.84 & 1 & 1.26 & 0.90 & &  & 1 & 1 & 1.99 \\
\hline\hline
8 & 4 & 2 & 1.13 & 1.41 & & &         & 1 & 0.75 & 0.31 & 0.75 & $-1.88$ \\
20 & 5 & 4 & 1.03 & 1.14 & 1.24 &  &  & 1 & 0.94 & 0.80  & 0.94 & 0.12 & $-1.85$ \\
%20 & 4 & 5 & 1.02 & 1.09 \\
32 & 4 & 8 & 1.008 & 1.04 & 1.08 &  &  & 1 & 0.98 & 0.95  & 0.98 & 0.77 & \\
%64 & 4 & 16 & 1.002  \\
%128 & 4 & 32 & 1.0005  \\
%\hline
\end{tabular}
\end{ruledtabular}
\end{table}

Table \ref{tableA} shows how well the coefficients $p_k,q_k,r_k$
of the small-$\xi$ expansion of the above parameters in the
dispersion laws are reproduced with $\mathbf v$-discretized
hard loops. In the case of regular polyhedra,
more and more coefficients $p_k,q_k,r_k$
become exact or come closer to the continuum result as $\mathcal N$ increases.
For the two pairs of polyhedra where the dispersion laws coincide,
there is particular good agreement with the continuum results
for the asymptotic transverse masses, however at the expense of having
one coefficient with a wrong sign in the longitudinal sector.

The lower part of Table \ref{tableA} performs the same\footnote{The
coefficients $q_{1,2}$ have been omitted out of laziness.}
comparison for a few cases of
the ``disco-ball'' discretization
(\ref{disco}) with
regular spacing in $z$ and $\varphi$. 
For $\mathcal N \le 20 $, the regular polyhedra 
turn out to superior in the small-$\xi$ expansion. However,
the disco-ball discretization allows us to increase $\mathcal N$
without limit, and while (with the exception of $r_1$) no coefficient
becomes exact, all of them can be systematically improved.
For example, choosing $N_z=16$ or 32 improves the coefficient
$p_1$ to become accurate to $0.2\%$ or $0.05\%$, respectively. 

The relation (\ref{mmvsmpl}) between the transverse plasma frequency and the
static mass parameter $\mu^2$ which determines 
the range of plasma instabilities
is in fact exact in all cases we have considered, so the coefficients $p_k$
in Table \ref{tableA} cover both the transverse plasma frequency and the
magnetic instability (resp.\ screening) mass scale for $\xi>0$ ($\xi<0$).

\section{Lattice equations of motion}
\label{applatt}

\subsection{1+1-dimensional case}

In the dimensionally reduced situation arising for
initial conditions depending only on $z$ and $t$,
we have adjoint scalar fields
$\phi_x=A_x$ and $\phi_y=A_y$ as well as auxiliary fields $\mathcal W_{\mathbf v}$ 
in the adjoint representation,
which live on the sites of the spatial lattice
with lattice spacing $a$. These are
represented as $N_c\times N_c$ traceless Hermitian matrices belonging
to the Lie algebra of SU$(N_c)$. 
Furthermore there are the conjugate momenta $\pi_x$ and $\pi_y$
(corresponding to $\partial_t A_x$ and $\partial_t A_y$) and 
$E_z=-\partial_t A_z$  which are all represented as $N_c\times N_c$ 
traceless Hermitian matrices living on the temporal links. 
Finally, there
are SU$(N_c)$ group elements $U$ which live on the spatial links.
The latter represent the discretized version of the parallel transporter
which in fundamental representation is given by
\beq
U(z^\prime,z)={\mathcal P}\exp{\left(- i g \int_{z}^{z^\prime}
d\ z^{\prime \prime} A_z (z^{\prime \prime})\right)}
\eeq
\comment{mpt}
and where ${\mathcal P}$ represents path ordering (with $z^\prime$
to the left and $z$ on the right). 
The discretized version then transports from the site $z$
to the site $z^{\prime}=z+a\rightarrow s+1$, so that
\beq
U_{s+\frac{1}{2}}=\exp{\left(- i g a A_{z,s}\right)}.
\eeq
\comment{mpt}
The covariant derivative $D_z$ may then be discretized into a 
right-transporting and a left-transporting version,
\beq
D_z^{R}\phi_s\rightarrow \frac{\phi_{s}-U_{s-\frac{1}{2}} \phi_{s-1} 
U_{s-\frac{1}{2}}^{\dagger}}{a}, \quad  D_z^{L}\phi_s\rightarrow %
\frac{U_{s+\frac{1}{2}}^{\dagger} \phi_{s+1} U_{s+\frac{1}{2}}-\phi_s}{a},
\eeq
\comment{mpt}
whereas the second order derivative is then symmetric since
\beq
D_z^2\rightarrow \frac{D_z^L-D_z^R}{a}.
\eeq
\comment{mpt}
Since the $E_{z,s}$ are living on the spatial links from $s$
to $s+1$, the first order covariant derivative has to be taken as $D_L$,
while the current is replaced by an average between the beginning and the
end of the link. Specifically,
\bqa
\pi_{\alpha,s}(t+\frac{\epsilon}{2})&=&\pi_{\alpha,s}(t-\frac{\epsilon}{2})+\epsilon \left[a^{-1}\left(%
D_z^{L}-D_z^{R}\right)\phi_s-g^2 [\phi_{\bar{\alpha},s},[\phi_{\bar{\alpha},s},
\phi_{\alpha,s}]]+j_{\alpha,s}\right]_t\nonumber\\
E_{z,s}(t+\frac{\epsilon}{2})&=&E_{z,s}(t-\frac{\epsilon}{2})+\epsilon\left[ig \sum_{\alpha}[\phi_{\alpha,s},D_z^L \phi_{\alpha,s}]-\frac{1}{2}\left(j_{z,s}+U_{s+\frac{1}{2}}^{\dagger} j_{z,s+1}%
U_{s+\frac{1}{2}}\right)\right]_t,\nonumber\\
\eqa
\comment{mpt}
where $\alpha$ runs over $x,y$ and $\bar{\alpha}$ denotes $y$ and $x$. 
The fields are then updated using symmetric first order spatial derivatives 
($D_z\rightarrow D_z^{S}=(D_z^{L}+D_{z}^{R})/2$),
\bqa
\phi_{\alpha,s}(t+\epsilon)&=&\phi_{\alpha,s}(t)+\epsilon \pi_{\alpha,s}(t+\frac{1}{2}\epsilon) \nonumber\\
{\mathcal W}_{{\mathbf v},s}(t+\epsilon)&=&
{\mathcal W}_{{\mathbf v},s}(t)+\epsilon \left[
-i g [{\mathcal A}(t),{\mathcal W}_{{\mathbf v}}(t)]+b_{\mathbf v} D_z^{S}
{\mathcal A}(t)-v_z D_z^{S} {\mathcal W}_{{\mathbf v},s}(t)-a_{\mathbf v} v_x \pi_x(t+\frac{\epsilon}{2})\right.\nonumber\\
&&\left.-a_{\mathbf v} v_y \pi_y(t+\frac{\epsilon}{2})\right]_{s}
+\epsilon%
\left[\frac{E_{z,s}(t+\frac{\epsilon}{2})+U_{s-\frac{1}{2}}(t) 
E_{z,s-1} (t+\frac{\epsilon}{2})
U_{s-\frac{1}{2}}^{\dagger}(t)}{2} (v_z a_{\mathbf v}+b_{\mathbf v})\right]
\nonumber\\
{\mathcal A}(t)&=&v_x \phi_x(t)+v_y \phi_y(t).
\eqa
\comment{pt}
Finally, choosing $E_{z,s}$ to transform at the spatial lattice
site $s$ rather than $s+1$ we find for the following update law for the
links $U$:
\beq
U_{s+\frac{1}{2}}(t+\epsilon)=U_{s+\frac{1}{2}}(t) 
\exp{\left(i g \epsilon a E_{z,s}(t+\frac{\epsilon}{2})\right)}.
\eeq
\comment{mpt}
Gauss's law becomes
\beq
i g \sum_{\alpha} [\phi_{\alpha,s}(t),
\pi_{\alpha,s}(t-\frac{\epsilon}{2})]-D_z^{R}E_{z,s}
(t-\frac{\epsilon}{2})+\frac{1}{N}\sum_{\mathbf v} {\mathcal W}_{{\mathbf v},s}
(t)=0
\eeq
\comment{mpt}
while the approximately conserved energy (\ref{Etot}) is discretized by
\bqa
{\mathcal E}&=&a\sum_{s} \Biggl\{\frac{1}{4}{\rm tr}\left[\left(E_{z,s}
(t-\frac{\epsilon}{2})+E_{z,s}(t+\frac{\epsilon}{2})\right)^2\right]+
\sum_{\alpha}
\Biggl[\frac{1}{4}{\rm tr}\left[\left(\pi_{\alpha,s}(t-\frac{\epsilon}{2})+
\pi_{\alpha,s}(t+\frac{\epsilon}{2})\right)^2\right]\nonumber\\
&&\left.\left.+{\rm tr}\left((D_z^{R}\phi_{\alpha,s}(t))^2\right)\right]+
{\rm tr}\left((i[\phi_{x,s},\phi_{y,s}])^2\right)_t+\epsilon%
\sum_{t^\prime=0}^{t} {\rm tr}\left[-\pi_{x,s}(t^\prime-\frac{\epsilon}{2})
\left(j_{x,s}(t^\prime-\epsilon)+j_{x,s}(t^\prime)\right)\right.\right.\nonumber\\
&&\left.\left.-\pi_{y,s}(t^\prime-\frac{\epsilon}{2})
\left(j_{y,s}(t^\prime-\epsilon)+j_{y,s}(t^\prime)\right)+
\frac{1}{2}\left(E_{z,s}(t^\prime-\frac{\epsilon}{2})+U_{s-\frac{1}{2}}(t^\prime)
E_{z,s-1}(t^\prime-\frac{\epsilon}{2})U^{\dagger}_{s-\frac{1}{2}}(t^\prime)
\right)\right.\right.\nonumber\\
&&\left(j_{z,s}(t^\prime-\epsilon)+j_{z,s}(t^\prime)\right)
\Biggr]\Biggr\},
\eqa
\comment{}
%where we used Refs.\textbf{insert ref to hep-ph/9401211 here} 
%form of the HL energy contribution and subsequently discretized it 
%appropriately.
As initial conditions we choose
\bqa\label{initcond1d}
&U_{s+\frac{1}{2}}(0)={\mathbf 1}_{N_c},\quad \phi_{\alpha,s}(0)=0,\quad 
E_{z,s}(-\frac{\epsilon}{2})=0, \quad {\mathcal W}_{{\mathbf v},s}(0)=0,&
\nonumber\\
&\langle \pi^a_{\alpha,s}(-\frac{\epsilon}{2}) \pi^b_{\beta,s^\prime}(
-\frac{\epsilon}{2}) \rangle
=\sigma^2 \delta^{ab} \delta_{\alpha \beta} \delta_{s s^\prime}/a\quad. &
\eqa
\comment{mpt}
%where $\sigma=0.01$.

Under local gauge transformations $\Lambda(z)$ one has
\bqa
\phi_{\alpha,s}\rightarrow \phi^{\prime}_{\alpha,s}&=&\Lambda_s \phi_{\alpha,s}
\Lambda^{-1}_s\nonumber\\
\pi_{\alpha,s}\rightarrow \pi^{\prime}_{\alpha,s}&=&\Lambda_s \pi_{\alpha,s}
\Lambda^{-1}_s\nonumber\\
E_{z,s}\rightarrow E^{\prime}_{z,s}&=&\Lambda_s E_{z,s}
\Lambda^{-1}_s\nonumber\\
{\mathcal W}_{{\mathbf v},s}\rightarrow {\mathcal W}_{{\mathbf v},s}^{\prime}
&=&\Lambda_s {\mathcal W}_{{\mathbf v},s} \Lambda^{-1}_s\nonumber\\
U_{s+\frac{1}{2}}\rightarrow U^{\prime}_{s+\frac{1}{2}}&=&\Lambda_{s+1} U_{s+\frac{1}{2}} \Lambda^{-1}_{s}
\eqa
\comment{p}
which can be used to check explicit gauge invariance of the results.

\subsection{3+1-dimensional case}

For the 3+1 dimensional simulation, we found it convenient to follow
very closely the lattice equations of motion used in Ref.~%
\cite{Bodeker:1999gx}. The dynamical variables are taken to be the three electric
fields $E_i(x)=-\partial_t A_i(x)$, lattice links $U_i(x)=\exp{(-i g a
A_i)}$ and the auxiliary fields $\mathcal{W}_{\bf v}(x)$.
Sticking to the convention of Ref.~\cite{Bodeker:1999gx}, this unfortunately creates a
slight inconsistency with the definition of the links $U_i$ in the
last subsection: for the three dimensional simulation, these are now
defined to transport from site $\vec{x}$ to site $\vec{x}- a
\hat{e}_i$. Furthermore, we make the field variables dimensionless by
absorbing the lattice spacing as well as the strong coupling constant
$g$ by 
\beq
g a A\rightarrow A,\qquad g a^a E\rightarrow E, \qquad g a^3
{\mathcal W}\rightarrow {\mathcal W}, \qquad g a^3 j\rightarrow j.
\eeq
Making use of the shorthand notation
${\bf x}+a {\bf \hat{e}}_i\ \rightarrow x+i$,
the lattice equations of motion then read for gauge group SU(2)
\bqa
E_{i}^a(t+\frac{\epsilon}{2},x)&=&E_{i}(t-\frac{\epsilon}{2},
x)+2i\epsilon\ {\rm tr}\left(
\tau^a U_{i}(t,x) \sum_{|j|\neq i}
S_{ij}^{\dagger}(t,x)\right)\nn\\
&&-\frac{\epsilon}{2}\left(j_i(t,x)+U_i(t,
x)j_i(t,x+i) U_i^{\dagger}(t,x)\right)\nn\\
U_{i}(t+\epsilon,x)&=&\exp{(i\ \epsilon\ E_{i}(t+\frac{\epsilon}{2},x))} U_{i}(t,x) \\
{\mathcal W}^a_{\bf v}(t+\epsilon,x)&=&{\mathcal W}^a_{\bf
v}(t,x)-\epsilon {\bf v}\cdot{\bf D}_{S}
{\mathcal W}^a_{\bf v}(t,x)+\epsilon \left( a_{\bf v} {\bf v} + b_{\bf v} \hat{\bf e}_z \right) \cdot{\bf E}^a(t+
\frac{\epsilon}{2})\nn\\
&&-2 i\epsilon b_{\bf v} \left(v_{x} {\rm tr}\left(\tau^a
U_{\Box,xz}(t,x)\right)+v_{y} {\rm tr}\left(\tau^a
U_{\Box,yz}(t,x)\right)\right),
\eqa
where
\beq
S_{ij}^{\dagger}(t,x)=U_j(t,x+i)U_i^{\dagger}(t,x+j)U_j^{\dagger}(t,x)
\eeq
is the gauge link staple, 
\beq
U_{\Box,ij}(t,x)=U_i(t,x)U_j(t,x+i)U^\dagger_i(x+j)U^\dagger_j(x)
\eeq
is the standard plaquette and the symmetric covariant derivative here is
now
\beq
D^S_{i}\phi(t,x)=\frac{1}{2}\left(U_i(t,x) \phi(t,x+i) U_i^\dagger(t,x)
  -U^\dagger_i(t,x-i) \phi(t,x-i) U_i(t,x-i)  \right).
\eeq 
Note that the sum
$\sum_{|j|\neq i}$ runs over both positive and negative directions,
with $U_{-j}(t,x)=U_{j}^\dagger(x-j)$.

Contrary to Ref.~\cite{Bodeker:1999gx} 
we did not discretize the first order temporal
derivatives in a symmetric way, thereby breaking explicit time
reversal symmetry which might introduce additional lattice discretization
artifacts in the simulation. However, by using different temporal step
sizes and different lattice spacings we have convinced ourselves that
our results are afflicted only very mildly by these artifacts
if we choose $\epsilon/a$ sufficiently small.

The energy density (\ref{Etot}) then reads
\bqa
{\mathcal E}&=&\frac{1}{g^2 a^4}\Biggl[4 \sum_{\Box}(1-\frac{1}{2}{\rm
tr}U_{\Box}(t,x))+\frac{1}{4}\sum_i {\rm tr} \left(E_i(t-\frac{\epsilon}{2},x)+
E_i(t+\frac{\epsilon}{2},x)\right)^2+\nn\\
&&\left.\epsilon \sum_i\sum_{t^\prime=0}^t {\rm
tr} E_i(t^\prime-\frac{\epsilon}{2},x)\left(j(t^\prime-\epsilon,x)+
j(t^\prime,x)+U_i(t^\prime,x)
j(t^\prime-\epsilon,x+i)U_i^\dagger(t^\prime,x)\right.\right.\nn\\
&&\left.
+U_i(t^\prime,x) j(t^\prime,x) U_i^\dagger(t^\prime,x)\right)\Biggr],
\eqa
and Gauss's law becomes
\bqa
\sum_i
\left(U_i^\dagger(t,x-i)E_i(t-\frac{\epsilon}{2},x-i)U_i(t,x-i)-
E_i(t-\frac{\epsilon}{2},x)\right)+\frac{1}{\mathcal N}\sum_{\bf v}
{\mathcal W}_{\bf v}(t,x)=0.
\eqa

Finally, we initialize our three-dimensional simulations with
$E_i(-\frac{\epsilon}{2},x)=A_i(0,x)=0$ and
\beq\label{3dinitcond}
\left<{\mathcal W}^a_{\bf v}(0,x) {\mathcal W}^b_{\bf v}(0,y)\right>=
\delta^{ab}\delta^3_{x,y}\sigma^2 a^3,
\eeq
where $\delta^3_{x,y}$ denotes 3 Kronecker deltas, and subtract
$\sum_{\bf v} {\mathcal W}^a/{\mathcal N}^2$ from each ${\mathcal
W^a}$  in order to obey
Gauss's law.

%\bibliography{bsample}
%\bibliographystyle{utphys}

%************************************************************************|
% Bibliography
%************************************************************************|

\end{document}